\documentclass[aps,pra,amsmath,amssymb,floatfix,twocolumn,amsmath,superscriptaddress,twocolumn,nofootinbib,tighten,letterpaper]{revtex4-2}
\usepackage[colorlinks,linkcolor=blue,citecolor=blue,urlcolor=blue]{hyperref}
\usepackage{multirow}
\usepackage{subfigure}
\usepackage{color}
\usepackage{mathrsfs}
\usepackage{hyperref}
\usepackage[normalem]{ulem}
\usepackage{bm}

\usepackage{amssymb}
\usepackage{amsmath}
\renewcommand\vec[1]{\ensuremath\boldsymbol{#1}}

\usepackage{tabularray}

\usepackage{array} 
\newcolumntype{P}[1]{>{\centering\arraybackslash}p{#1}}

\definecolor{RowColor}{rgb}{0.88,1,0.9}

\usepackage{amsfonts, relsize, color, mathtools, physics}
\usepackage{graphics}
\usepackage{graphicx}
\usepackage{subfigure}
\usepackage{color}
\usepackage{comment}

\begin{document}

\title{Amplified magnetic catalysis in non-Hermitian Euclidean and hyperbolic Dirac liquids}

\author{Christopher A. Leong}
\affiliation{Department of Physics, Lehigh University, Bethlehem, Pennsylvania, 18015, USA}

\author{Bitan Roy}
\affiliation{Department of Physics, Lehigh University, Bethlehem, Pennsylvania, 18015, USA}

\date{\today}

\begin{abstract}
Due to their iconic linearly vanishing density of states near the zero-energy, half-filled two-dimensional Dirac materials in flat Euclidean and negatively-curved hyperbolic spaces exhibit dynamic mass generation only once a critical interaction strength is surpassed. Application of sufficiently strong perpendicular external magnetic fields (uniform or inhomogeneous) onto these systems can, however, trigger the formation of such ordered phases yielding isotropic insulation near the band-center at arbitrarily weak coupling, a phenomenon driven by the topologically robust manifold of zero-energy states and known as magnetic catalysis. Recently, it has been proposed that a specific type of non-Hermiticity, allowing the system to feature an all-real eigenvalue spectrum otherwise squeezed toward the zero-energy, can bring down the requisite critical coupling of a specific family of ordered phases, named as commuting class masses, to a desired lower finite value in Dirac systems, a phenomenon known as non-Hermitian catalysis (\href{https://arxiv.org/abs/2501.18591}{arXiv:2501.18591}). Here, from concrete symmetry-based arguments, we predict that a confluence of external magnetic fields and such a non-Hermiticity can amplify the magnitude of commuting class masses for subcritical strengths of interactions in Dirac liquids, an intriguing emergent phenomenon hereafter named non-Hermitian amplification of magnetic catalysis. We anchor this prediction from numerical self-consistent mean-field solutions of the charge-density-wave (antiferromagnetic) order displaying a staggered pattern of average electronic density (magnetization) between the nearest neighboring sites of the half-filled Euclidean honeycomb and hyperbolic $\{ 10, 3\}$ and $\{ 14, 3\}$ (Schl\"afli symbols) lattices, all featuring emergent non-Hermitian Dirac quasiparticles, after decomposing the nearest-neighbor Coulomb (on-site Hubbard) repulsion in the Hartree channel and by following general principles of biorthogonal quantum mechanics. With the non-Hermiticity stemming from an imbalance in the hopping amplitudes between the nearest-neighbor sites in opposite directions, both of the orders belong to the commuting class mass family. We discuss the scaling behavior of these two orders with the strength of magnetic field and non-Hermiticity in these systems over a wide range of subcritical interactions. Possible table-top experimental setups to test our predictions are discussed.          
\end{abstract}

\maketitle

\section{Introduction}~\label{sec:introduction}

A two-dimensional electron gas (2DEG), typically realized in GaAs heterostructures, for example, when placed in strong perpendicular magnetic fields, features many complex phenomena, among which quantum oscillations~\cite{QOcillation:1, QOcillation:2}, and integer and fractional quantum Hall effects~\cite{QHE:book} are the most prominent ones. An atom-thick single-layer honeycomb lattice of carbon atoms, known as graphene, when placed in strong magnetic fields also displays quantized Hall plateaus at integer fillings, bearing the signature of relativistic quasiparticles, confined to a flat Euclidean plane~\cite{grapheneQH:1, grapheneQH:2, grapheneQH:3}. What makes relativistic 2DEG distinct from their conventional (non-relativistic) counterparts is the existence of topologically robust zeroth Landau level that survives even when the magnetic field assumes a spatially modulated profile as long as the system encloses a finite number of magnetic flux quanta~\cite{aharonovcasher} and the filled sea of Dirac Landau levels at negative energies in half-filled systems.

Together these two ingredients allow the formation of Dirac masses at and near the half-filling via spontaneous breaking of discrete and/or continuous symmetries for infinitesimal strength of Hubbard-like local interactions, for example, yielding an isotropic insulation within the zeroth Landau level, a phenomenon known as \emph{magnetic catalysis}~\cite{magcata:1, magcata:2, magcata:3, magcata:4, magcata:5}. This mechanism is responsible for the formation of the quantum Hall states at integer fillings of $\nu=0$ and $\pm 1$ within the manifold of spin and valley degenerate, but sublattice-polarized zeroth Landau levels in graphene~\cite{magcatagraphene:1, magcatagraphene:2, magcatagraphene:3, magcatagraphene:4, magcatagraphene:5, magcatagraphene:6, magcatagraphene:7}, observed in experiments~\cite{graQH01:1, graQH01:2, graQH01:3, graQH01:4}. Recently, this mechanism has also been established for two-dimensional hyperbolic Dirac systems subject to either uniform or inhomogeneous magnetic fields~\cite{Bfieldhyperbolic:3}, otherwise featuring the characteristic linearly vanishing density of states (DOS) near the zero-energy in pristine condition~\cite{hyperbolicclassification}. Furthermore, magnetic catalysis can give rise to axionic insulators at weak coupling in three-dimensional Weyl semimetals~\cite{magcataWeyl:1, magcataWeyl:2, magcataWeyl:3, magcataWeyl:4, magcataWeyl:5}. In all these cases, due to the presence of external magnetic fields, the fermionic sector dynamically acquires mass via spontaneous symmetry breaking in the spirit of the Anderson-Higgs mechanism at arbitrarily weak coupling~\cite{Peskin, ZinnJustin}.

\begin{figure*}[t!]
    \includegraphics[width=1.0\linewidth]{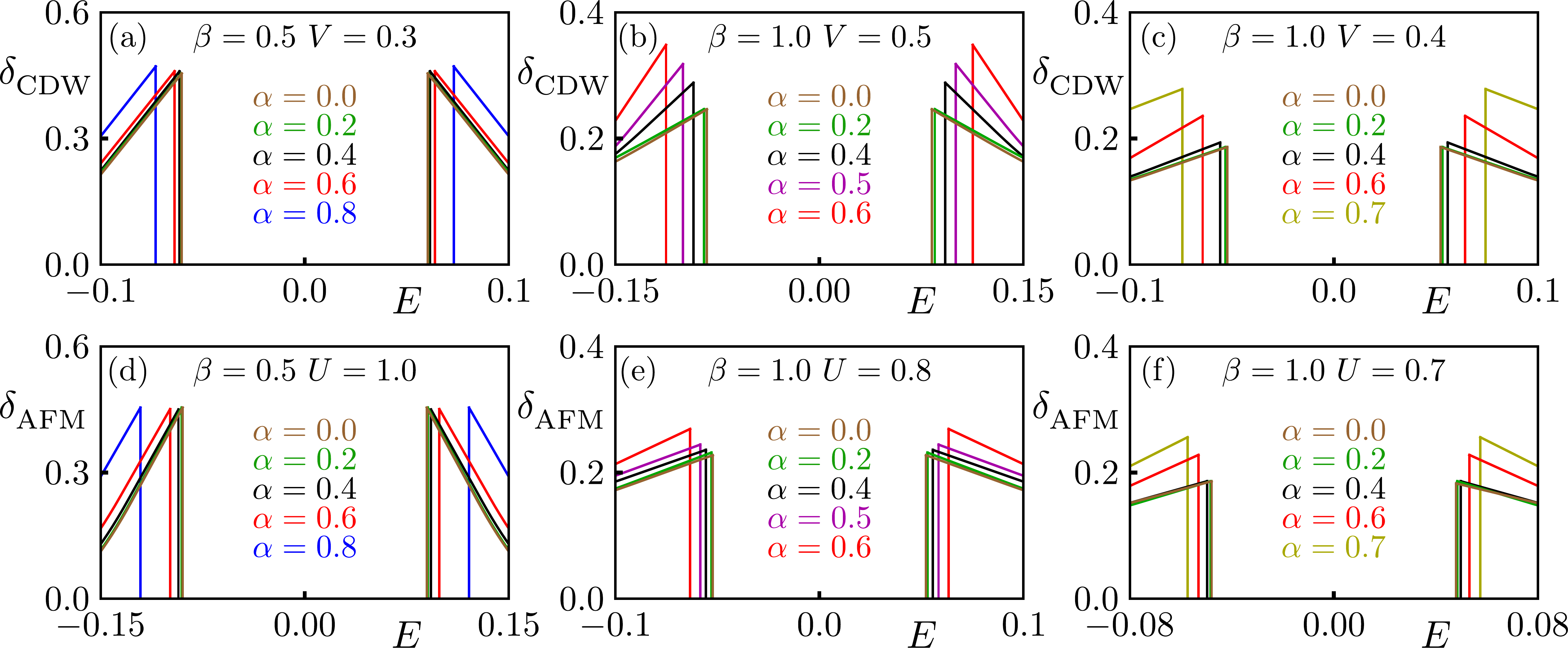}
    \caption{Density of states (DOS) over a small energy ($E$) window around $E=0$ in Hermitian ($\alpha=0$) and non-Hermitian (finite $\alpha$) (a) Euclidean $\{6, 3\}$ or honeycomb, (b) hyperbolic $\{10, 3\}$, and (c) hyperbolic $\{14, 3\}$ lattices at a fixed magnetic field flux per plaquette $\beta$ (measured in units of $\Phi_0/(2 \pi)$, where $\Phi_0$ is the magnetic flux quanta) and subcritical (with respect to the zero magnetic field case) nearest-neighbor (NN) Coulomb repulsion $V$ (see Table~\ref{tab:citicalvalues}). The results are obtained from the self-consistent solutions of the charge-density-wave (CDW) order, a member of the commuting class mass family in our construction of non-Hermitian Dirac system, where non-Hermiticity stems from an imbalance of the hopping amplitudes between the NN sites in the opposite directions, see Fig.~\ref{fig:NHpeierlssub}. Panels (d)-(f) are analogous to (a)-(c), respectively, but DOS in all these systems are obtained from the self-consistent solutions of the antiferromagnet (AFM) order, yet another member of the commuting class mass family in our specific construction of non-Hermitian Dirac systems, in the presence of subcritical on-site Hubbard repulsion $U$ (see Table~\ref{tab:citicalvalues}). The formation of the spectral gaps for subcritical $V$ and $U$ in Hermitian systems confirms their magnetic catalysis, which remains operative in non-Hermitian systems. Results show that the spectral gap near the zero-energy increases with increasing non-Hermiticity for subcritical $V$ and $U$ in all these systems  when they are subject to uniform external magnetic fields, thereby endorsing our proposed scenario of non-Hermitian amplification of magnetic catalysis. See legends for specifications of various parameter values.      
    }~\label{fig:Summary}
\end{figure*}

Recently, an alternative route has been proposed to bring the requisite critical interaction strength for dynamic mass generation in Dirac systems down to a desired lower coupling by exposing them to a decorated environment, sourcing balanced gain and loss to accommodate a non-Hermitian (NH) platform, a phenomenon we named \emph{non-Hermitian catalysis}~\cite{NHcatalysis:1}. This mechanism rests on the key facts that the corresponding NH operators continue to display the hallmark linearly vanishing DOS in two-dimensional NH Dirac systems, manifesting an all-real eigenvalue spectrum over an extended parameter regime, while the bandwidth gets progressively reduced with increasing non-Hermiticity in the system, quantifying its coupling with the environment. Over this parameter regime, thus all the single-particle states are sharp (possessing an infinite lifetime) and the filling-factor remains well defined. In such NH or open Dirac systems, a subset of Dirac masses, called \emph{commuting class masses} (defined later) can be nucleated for weaker interactions in comparison to their counterpart in closed or Hermitian systems, which can be tuned desirably by controlling the degree of non-Hermiticity in the system. In this work, we set out to showcase an intriguing confluence of these two catalysis mechanisms (magnetic and non-Hermitian), leading to an amplified magnetic catalysis of commuting class masses triggered by the non-Hermiticity in the system, shown in in Fig.~\ref{fig:Summary}. In this context, our main findings are summarized next.

\subsection{Key Results}

\begin{figure}[t!]
    \includegraphics[width=1.0\linewidth]{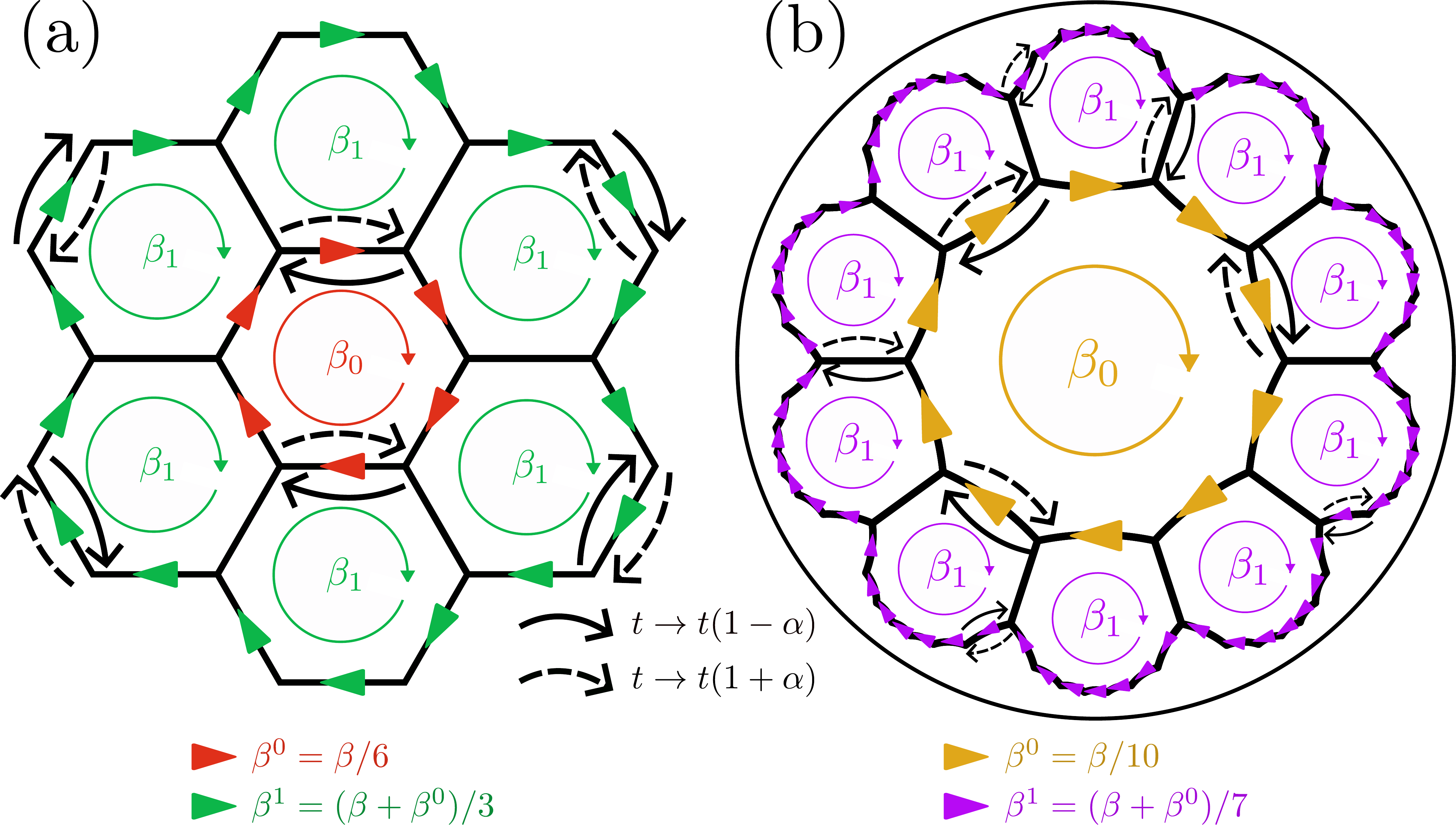}
    \caption{Schematic representations of a uniform magnetic flux attachment in non-Hermitian (NH) Dirac systems on (a) Euclidean $\{6, 3\}$ or honeycomb and (b) hyperbolic $\{10, 3\}$ lattices, in which the non-Hermiticity, quantified by $\alpha$, resulting from an imbalance in the hopping amplitudes between the nearest-neighbor (NN) sites in the opposite directions, given by $t(1 \pm \alpha)$, is represented by solid and dashed curved black arrows in their directions. Here, $t$ is the reciprocal hopping amplitude and $\alpha$ measures the non-Hermiticity. The colored arrow head represents the Peierls phase added to each NN hopping term in its direction, yielding a uniform perpendicular external magnetic field in NH systems, following Eq.~\eqref{eq:NHGeneral}. Notice that the phases are threaded around each plaquette in a way such that the total phase accumulated in the clockwise direction around each plaquette is always equal to $\beta$, yielding the magnetic flux [Eq.~\eqref{eq:psmagflux}]. The black circle in (b) corresponds to the conformal boundary of the Poincar\'e disk representation. With $\beta^0=\beta/14$ and $\beta^1=(\beta + \beta^0)/11$, the construction from (b) can immediately be generalized to $\{ 14,3 \}$ hyperbolic lattice to mimic a uniform magnetic field therein. 
    }~\label{fig:NHpeierlssub}
\end{figure}

In this work, we consider Euclidean $\{6,3\}$ or honeycomb and hyperbolic $\{ 10,3 \}$ and $\{ 14,3 \}$ lattices, featuring emergent gapless Dirac fermions on flat and negatively curved spaces, respectively, resulting from a simple nearest-neighbor (NN) tight-binding model with a spin-independent hopping amplitude therein~\cite{grpahene:Wallace, grpahene:Semenoff, graphene:RMP, hyperbolicclassification}. Here, $\{ p, q\}$ is the Schl\"afli symbol, defined in Sec.~\ref{sec:system}. The orbital coupling of the external magnetic fields with fermions is captured by attaching the Peierls phase to the NN hopping matrix elements. The non-Hermiticity in our construction results from a real-valued imbalance of the hopping amplitudes between two sublattices of all these bipartite lattices in the opposite directions, as shown in Fig.~\ref{fig:NHpeierlssub}. Such a NH generalization of Dirac systems continues to show all-real eigenvalues over an extended parameter regime, on which we exclusively focus on in this work. Firstly, we note that the zero-energy manifold, responsible for magnetic catalysis of Dirac masses at half-filling in Hermitian systems, remains unaffected in the presence of non-Hermiticity in the system. This observation, consistent with recent predictions from continuum theory~\cite{NHAharonovCashed}, strongly suggests that the magnetic catalysis mechanism should remain operative in weakly correlated NH Dirac systems.

Due to the bipartite nature of all the lattice-regularized Dirac systems, the charge-density-wave (CDW) and antiferromagnet (AFM) orders are two natural candidates for Dirac masses that yield isotropic insulation at half-filling. Respectively, these two orders display staggered patterns of electronic density and magnetization between two sublattices, and are favored by NN Coulomb ($V$) and on-site Hubbard ($U$) repulsions. Owing to the vanishing DOS, nucleation of any such order takes place beyond a critical strength of the corresponding interaction in pristine Hermitian and NH Dirac systems (see Table~\ref{tab:citicalvalues}). However, the presence of external magnetic fields, by virtue of supporting a robust zero-energy manifold, triggers the nucleation of these two orders even for infinitesimal $V$ or $U$ (magnetic catalysis).

In this work, we show that a specific class of non-Hermiticity that squeezes all the states at finite energies closer to the band-center while leaving the zero-energy manifold unaffected, thereby reducing the corresponding bandwidth, boosts the magnetic catalysis of CDW and AFM orders by amplifying the magnitude of the corresponding order parameter. Such an outcome rests on the fact that these two orders fully anticommute with the total NH operator and thus stand as mass orders for the collection of NH Dirac fermions. This conclusion can be confirmed by comparing the gap magnitudes in the DOS near zero-energy in Hermitian and NH systems, as shown in Fig.~\ref{fig:Summary}. The reconstructed DOS is computed from the lattice-based numerical self-consistent solutions of CDW and AFM order parameters obtained by employing some key concepts of bi-orthogonal quantum mechanics for subcritical strengths of the NN Coulomb and on-site Hubbard repulsions after decomposing them in the Hartree channel. Namely, we find that the magnitude of the insulating gap with the CDW and AFM orders (belonging to the commuting class mass family in our construction of NH Dirac systems) increases monotonically with increasing non-Hermiticity in the system for a fixed strength of the external magnetic field and subcritical interaction ($V<V_c$ and $U<U_c$), endorsing the titled claim of this work, amplified magnetic catalysis in Euclidean and hyperbolic non-Hermitian Dirac liquids.

\begin{table}[t!]
\centering
{\renewcommand{\arraystretch}{1.2}
\begin{tabular}{|P{0.5cm}|P{1cm}|P{1cm}|P{1cm}||P{1cm}|P{1cm}|P{1cm}|}
        \hline
        \multicolumn{1}{|c|}{} & \multicolumn{3}{|c||}{Critical NN Coulomb} & \multicolumn{3}{|c|}{Critical on-site Hubbard}\\
        \cline{2-7}
        \vspace{-0.4cm}\hspace{0.1cm}$\alpha$ & $\{6, 3\}$ & $\{10, 3\}$ & $\{14, 3\}$ & $\{6, 3\}$ & $\{10, 3\}$ & $\{14, 3\}$\\
        \hline
        0.0 & 0.695 & 0.670 & 0.652 & 2.08 & 1.46 & 1.34\\
        0.2 & 0.680 & 0.658 & 0.636 & 2.06 & 1.43 & 1.32 \\ 
        0.4 & 0.637 & 0.617 & 0.595 & 1.91 & 1.34 & 1.24\\
        0.5 & $-$ & 0.585 & 0.565 & $-$ & 1.28 & 1.20 \\ 
        0.6 & 0.562 & 0.541 & 0.517 & 1.68 & 1.16 & 1.08 \\ 
        0.7 & $-$ & $-$ & 0.460 & $-$ & $-$ & 0.97 \\ 
        0.8 & 0.420 & $-$ & 0.387 & 1.24 & $-$ & 0.79 \\
        \hline
\end{tabular}
}
\caption{Critical values of the repulsive nearest-neighbor (NN) Coulomb and on-site Hubbard interactions on Euclidean honeycomb or $\{6, 3\}$, hyperbolic $\{10, 3\}$, and hyperbolic $\{14, 3\}$ lattices with varying degree of non-Hermiticity ($\alpha$) that also includes the Hermitian system ($\alpha=0$), in the absence of any external magnetic field~\cite{NHcatalysis:1}. Here, we picked the values of $\alpha$ that are being used in our numerical simulations in various systems. Dashed lines for specific values of $\alpha$ in a specific system indicate that those values of $\alpha$ are not being used in that system. 
}~\label{tab:citicalvalues}
\end{table}

\subsection{Organization}

The rest of the paper is organized as follows. In Sec.~\ref{sec:system}, we specify the requisite details (number of lattice sites and boundary conditions) of all the lattice-based Dirac systems we investigate in this work. Free-fermion tight-binding operators describing a collection of planar NH Dirac fermions in Euclidean and hyperbolic spaces, featuring an all-real eigenvalue spectrum over an extended parameter regime, in the presence of external perpendicular magnetic fields is introduced in Sec.~\ref{sec:nhfree}. The resulting robust manifold of zero-energy states and the scaling of the associated DOS in these systems are discussed in Sec.~\ref{sec:zeromodes}. The concept of NH amplification of magnetic catalysis for commuting class masses, supported by analytical arguments is promoted in Sec.~\ref{sec:NHamplMagCata}. Necessary tools of bi-orthogonal quantum mechanics are briefly reviewed in Sec.~\ref{sec:biorthogonal}. Numerical self-consistent solutions for the CDW and AFM orders in the mean-field approximation following the Hartree decompositions of the NN Coulomb and on-site Hubbard interactions for spinless and spinful fermions are discussed in Sec.~\ref{sec:coulomb} and Sec.~\ref{sec:hubbard}, respectively. Finite size scaling analysis of both the orders in systems with open boundary conditions is presented in Sec.~\ref{sec:finitesize}. Section~\ref{sec:summary} is devoted to concluding remarks and future directions related to this work, as well as to discussions on possible table-top experimental arrangements to test our predictions.

\section{Background and system specifications}~\label{sec:system}

Lattices are formed from periodic arrangement of regular polygons in a translationally invariant fashion that, depending on the connectivity of each site, features various discrete rotational symmetries. When such a polygon is composed of $p$ arms and each vertex is connected to $q$ neighboring ones, we can characterize the emergent lattice structure using the Schl\"afli symbol $\{ p, q\}$. On the two-dimensional flat Euclidean plane such a periodic arrangement can be accomplished only when the following condition is satisfied $(p-2) (q-2)=4$, which has only three solutions, corresponding to a honeycomb lattice with $p=2q=6$, a square lattice with $p=q=4$, and a triangular lattice with $2p=q=6$. Only the foremost system yields massless Dirac fermions as emergent quasiparticles near the half-filling, which can be demonstrated from a simple NN tight-binding model therein~\cite{grpahene:Wallace, grpahene:Semenoff, graphene:RMP}. By contrast, an infinite number of periodic tessellation can be accomplished on a hyperbolic plane with a constant negative curvature as then only an inequality $(p-2)(q-2) > 4$ needs to be satisfied. Among them a family of systems in which $q=3$ and $p/2$ is an odd integer harbor gapless Dirac-like excitations with the iconic linearly vanishing DOS near the zero-energy~\cite{hyperbolicclassification}. In this work, we consider the Euclidean $\{ 6,3 \}$ or honeycomb and $\{ 10, 3 \}$ and $\{ 14,3\}$ hyperbolic lattices. Hyperbolic lattices, in particular, often feature many peculiar outcomes solely arising due to the background constant negative curvature that have recently sparked a series of investigations~\cite{HL:1, HL:2, HL:3, HL:4, HL:5, HL:6, HL:7, HL:8, HL:9, HL:10}.

As all the results presented in this work are obtained from numerical calculations, it is worth announcing the system sizes in terms of the number of lattice points and the boundary conditions we impose therein. Throughout we work with these lattice-regularized Dirac systems with open boundary conditions. We also characterize the lattice systems in terms of their total generation number, defined in the following way. The central $p$-gon belongs to the zeroth generation ($n=0$) and each successive layer of plaquettes constitutes its progressively next generation of all these systems. Unless otherwise mentioned, throughout we work with honeycomb lattices containing nineteen generations and composed of 2400 sites, and  $\{10, 3\}$ ($\{14, 3\}$) hyperbolic lattices with three (two) generations, containing 2880 (1694) sites. Here, we always impose open boundary conditions. As the number of lattice sites increases very rapidly with the generation number in hyperbolic lattices, we cannot access such systems with any higher generation number.

\section{Dirac fermions in magnetic fields}~\label{sec:nhfree}

We begin the discussion with the tight-binding model for free fermions ($H_0$) in Euclidean $\{ 6,3 \}$ or honeycomb and hyperbolic $\{ 10,3 \}$ and $\{ 14,3 \}$ lattices with only the NN hopping amplitude (all yielding massless Dirac fermions near the half-filling in pristine condition), when external perpendicular magnetic fields pierce these systems. As all these lattice systems result from periodic arrangement of $p$-gons (regular polygons with $p$ arms) with even integer $p$, the NN tight-binding model always manifests a bipartite lattice structure (microscopic or emergent). Then the NN sites are identified to belong to two sublattices $A$ and $B$, although such labeling is arbitrary, unfolding a sublattice exchange symmetry of these systems. The orbital effect of the external perpendicular magnetic fields is captured by the Peierls substitution, resulting in a phase (Peierls phase) attached to the NN hopping amplitude~\cite{Peierls}. The corresponding tight-binding Hamiltonian for a collection of spinless fermions reads as 
\begin{equation}~\label{eq:TBPeierls}
H_0 =-\sum_{j \in A} \sum_{k \in B}{}^{'} t_{jk}\; c^\dagger_j \; \exp[i \beta_{jk}] \; c_k + {\rm H.c.}\;,  
\end{equation}    
where $i=\sqrt{-1}$, $c^\dagger_j$ ($c_j$) is the fermionic creation (annihilation) operator on the $j$th site, the prime symbol restricts the summation within the NN sites, and $\beta_{jk}$ is the Peierls phase between the $j$th and $k$th sites. The spin independent NN hopping amplitude $t_{jk}$ is assumed to be constant $t$, which we set to be unity. Here for simplicity we neglect the spin degrees of freedom as its inclusion leads to a mere doubling of the Hamiltonian since the hopping amplitude is spin-independent and throughout we neglect the Zeeman coupling between electronic spin and external magnetic fields. This prescription can be employed on all the lattices considered here to mimic any particular spatial profile of the magnetic fields. But, for the sake of concreteness, we restrict ourselves to uniform magnetic fields in this work, as shown in Fig.~\ref{fig:NHpeierlssub}.

\begin{figure*}
    \includegraphics[width=\linewidth]{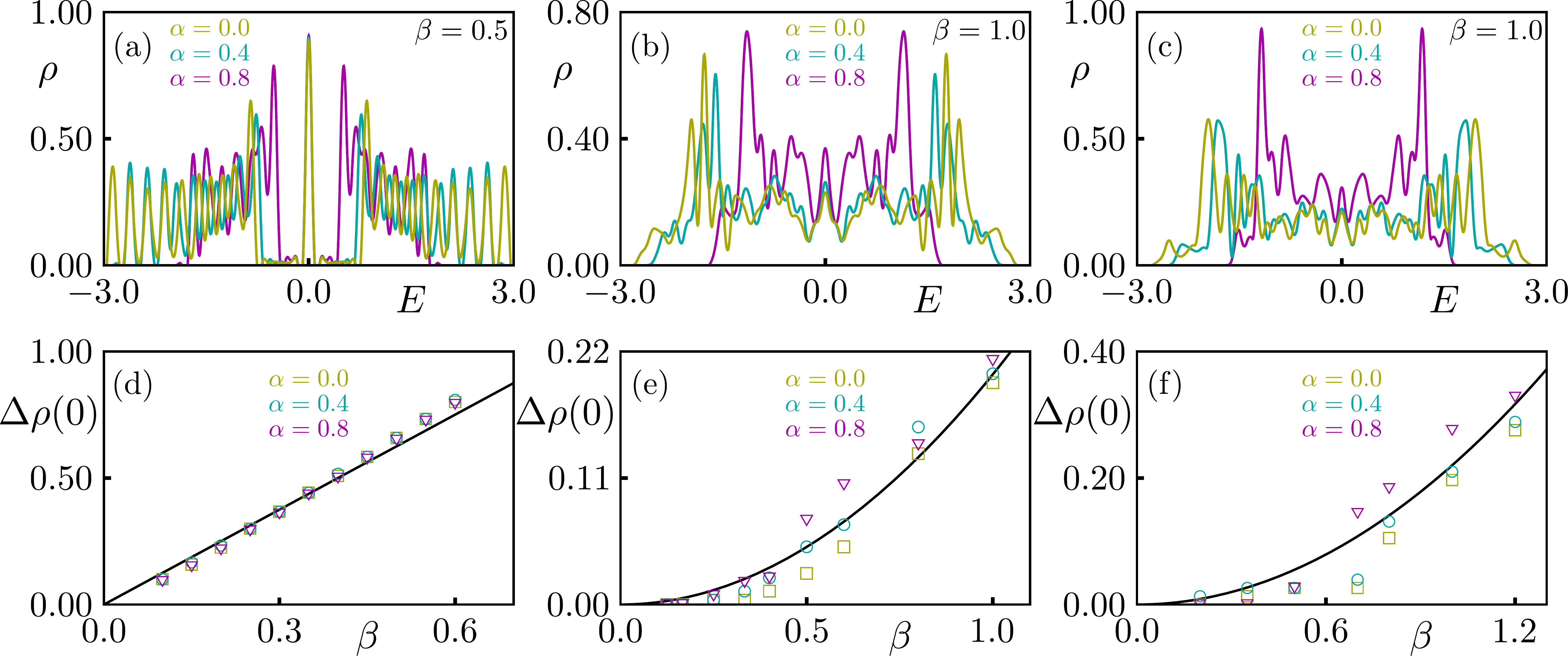}
    \caption{Density of states (DOS) $\rho$ as a function of energy in (a) Euclidean $\{6, 3\}$ or honeycomb, (b) hyperbolic $\{10, 3\}$, and (c) hyperbolic $\{14, 3\}$ lattices with a fixed non-zero total magnetic flux ($\beta$) in Hermitian ($\alpha=0$) and non-Hermitian (finite $\alpha$) Dirac systems. For $\alpha=0$, but with a finite and uniform $\beta$, a large (finite) number of near zero energy states emerges on honeycomb (hyperbolic) lattice(s), in agreement with an index theorem. While on honeycomb lattice we observe a clear formation of a few discrete Landau levels at finite energies, any such Landau quantization is absent on hyperbolic lattices due to the background constant negative curvature. Non-Hermiticity ($\alpha$) does not change the number of exact zero-energy modes, but it squeezes all the states at finite energies closer to the zero energy. The scaling of the near zero-energy DOS, defined as $\Delta \rho(0) = \rho_{\beta}(E=0) - \rho_{0}(E=0)$, where $\rho_{\beta}(E=0)$ [$\rho_{0}(E=0)$] is the near zero-energy DOS with [without] magnetic flux, in (d) $\{ 6,3 \}$, (e) $\{ 10,3 \}$, and (f) $\{ 14,3 \}$ lattices with $\beta$ for a few choices of $\alpha$. Notice that the subtraction of such a zero-field DOS, $\rho_{0}(E=0)$, is necessary to avoid systematic error brought on by the finite-width of the zero-energy bin. On the honeycomb lattice $\Delta \rho(0)$ scales linearly with $\beta$ and the numerically computed slope of such a linear fit is insensitive to $\alpha$ due to a finite Landau gap in the spectrum, which are consistent with the predictions of the Aharonov-Casher index theorem in Hermitian and non-Hermitian Euclidean Dirac systems. On the hyperbolic lattices $\Delta \rho(0)$ scales roughly as $\beta^2$ and a slight enhancement in $\Delta \rho(0)$ with finite $\alpha$ in our numerical calculation results from the absence of any Landau gap in the spectrum, causing minor deviation from the $\alpha$-independent $\beta^2$ scaling of $\Delta \rho(0)$ in non-Hermitian systems. 
    }~\label{fig:2_DOSFixedBVaryA}
\end{figure*}

With the prescription shown in Fig.~\ref{fig:NHpeierlssub}, the total Peierls phase ($\beta$) accumulated by traversing around any plaquette in the clockwise direction (chosen as a convention) is always given by
\begin{equation}
    \beta = 2\pi \; \left( \Phi_B/\Phi_0 \right),
    \label{eq:psmagflux}
\end{equation}
where $\Phi_B$ is the resulting \emph{uniform} magnetic flux through each plaquette and $\Phi_0$ is the magnetic flux quanta. Therefore, $\beta$ is the total magnetic flux through each plaquette in units of $\Phi_0/(2 \pi)$ and the total magnetic flux enclosed by the entire system is $\Phi_B$ times the total number of plaquettes in the system. From hereon, we refer to $\beta$ as the strength of the uniform magnetic field. With the underlying bipartite lattice structure, the Hamiltonian operator resulting from the NN tight-binding model [Eq.~\eqref{eq:TBPeierls}] in the presence of external magnetic field reads  
\begin{equation}~\label{eq:hamiltBfield}
\hat{h}_0 = \left(
\begin{array}{cc}
{\boldsymbol 0} & {\boldsymbol t}_{\beta}\\
{\boldsymbol t}^\dagger_{\beta} & {\boldsymbol 0}
\end{array}
\right),
\end{equation}
where ${\boldsymbol{t_{\beta}}}$ is the $N/2$-dimensional inter-sublattice matrix component of $H_0$, which depends on the magnetic field strength $\beta$, `$\dagger$' denotes the Hermitian conjugate, and ${\boldsymbol 0}$ is an $N/2$-dimensional null matrix. The corresponding spinor basis is $\Psi^\top = \left( c_A , c_B \right)$, where $c_A$ ($c_B$) is an $N/2$-dimensional spinor, constituted by the annihilation operators on the sites of the sublattice $A$ ($B$).

Construction of a NH generation of this scenario in which the corresponding NH operator features a guaranteed all-real eigenvalue spectrum over an extended NH parameter regime rests on the identification of a operator $\hat{h}_{\rm mass}$ that fully anticommutes with $\hat{h}_0$. Due to the bipartite nature of all the lattices yielding gapless emergent Dirac quasiparticles near the half-filling, we immediately identify one candidate for $\hat{h}_{\rm mass}$, given by 
\begin{equation}~\label{eq:hmass}
\hat{h}_{\rm mass} =  {\rm diag}. \left( {\boldsymbol I}, - {\boldsymbol I} \right),
\end{equation}
where ${\boldsymbol I}$ is an $N/2$-dimensional identity matrix. Due to the anticommutation relation $\{ \hat{h}_0, \hat{h}_{\rm mass} \}=0$, such a matrix operator is also called the mass matrix. With these ingredients in hand, we propose the following form of a NH operator~\cite{NHDirac:1, NHDirac:2, NHDirac:3, NHDirac:4, NHDirac:5}
\begin{equation}~\label{eq:NHGeneral}
    \hat{h}_{\rm NH} = \hat{h}_0 + \alpha \; \hat{h}_{\rm mass} \; \hat{h}_0
    = \left(
\begin{array}{cc}
{\boldsymbol 0} & (1 + \alpha){\boldsymbol{t_{\beta}}}\\
(1 - \alpha){\boldsymbol{t^{\dagger}_{\beta}}} & {\boldsymbol {0}}
\end{array}
\right),
\end{equation}
where $\alpha$ is a real-valued parameter that controls the non-Hermiticity in the system, with $\alpha=0$ corresponding to Hermitian systems. The non-reciprocal hopping patterns associated with the above NH operator are shown in Fig.~\ref{fig:NHpeierlssub} on $\{ 6,3 \}$ and $\{ 10, 3 \}$ lattices.

Finally, we comment on the eigenspectrum of $\hat{h}_{\rm NH}$. Since $\{ \hat{h}_0, \hat{h}_{\rm mass} \}=0$, the eigenspectrum of $\hat{h}_0$ is particle-hole symmetric, as $\hat{h}_{\rm mass}$ generates its unitary particle-hole or chiral or sublattice symmetry. Let us assume that the eigenvalues of $\hat{h}_0$ constitute a set $\pm \{ E_i \}$. Notice that $\{ \hat{h}_{\rm mass}, \hat{h}_{\rm NH} \}=0$ and therefore the spectrum of $\hat{h}_{\rm NH}$ is also particle-hole symmetric. As the anti-Hermitian component of $\hat{h}_{\rm NH}$, namely $\hat{h}_{\rm mass}\hat{h}_0$ anti-commutes with $\hat{h}_0$, the eigenvalues of $\hat{h}_{\rm NH}$ constitute a set given by $\pm \sqrt{1-\alpha^2} \{ E_i \}$. Notice that all the eigenvalues $E_i$ are functions of magnetic fields $\beta$, which we do not display explicitly here, unless there are states at precise zero energy. Therefore, the spectrum of $\hat{h}_{\rm NH}$ is composed of all-real eigenvalues for $|\alpha|<1$ and throughout we restrict ourselves within this parameter regime. Also notice that the scaling of the DOS does not get affected by the non-Hermiticity in the system, although the entire spectrum gets squeezed toward zero-energy with increasing $\alpha$. Thus, in the absence of external magnetic fields, the hallmark linearly vanishing DOS near zero-energy for Dirac liquids remains unchanged for any $|\alpha|<1$, thereby yielding NH Dirac liquids. In this construction, $\alpha=\pm 1$ mark the exceptional points, where all the eigenvalues of $\hat{h}_{\rm NH}$ coalesce to zero. In this work, we do not consider such singular points.

With the inclusion of the spin degrees of freedom, the NH operator for free-fermion systems reads as $\sigma_0 \otimes \hat{h}_{\rm NH}$. Here, $\{ \sigma_\nu \}$ is the set of Pauli matrices, operating on the spin index, with $\nu=0,1,2,3$ and $\otimes$ is the Kronecker product.

\section{Zero Modes in NH Dirac systems}~\label{sec:zeromodes}

In the previous section, we have presented the construction of lattice-regularized NH Dirac models subject to uniform external magnetic fields with an all-real eigenspectrum over an extended NH parameter regime ($|\alpha|<1$), see Eq.~\eqref{eq:NHGeneral}. Before delving into the effects of electronic interactions in such systems, we numerically study the behavior of these models in terms of the scaling of the DOS in the presence of an external magnetic field and non-Hermiticity in the system. To this end, we numerically diagonalize $\hat{h}_{\rm NH}$ for Euclidean $\{6, 3\}$ or honeycomb, and hyperbolic $\{10, 3\}$ and $\{14, 3\}$ lattices. In particular, we are interested in the scaling of the DOS with $\beta$ and $\alpha$ with a special focus near the zero-energy. The DOS at a given energy $E$ is obtained by counting the number of states within an energy bin of a small width ($\delta E$), centered around $E$. Results of this analysis are shown in Fig.~\ref{fig:2_DOSFixedBVaryA}, which we discuss next.

\begin{figure*}[t!]
    \includegraphics[width=\linewidth]{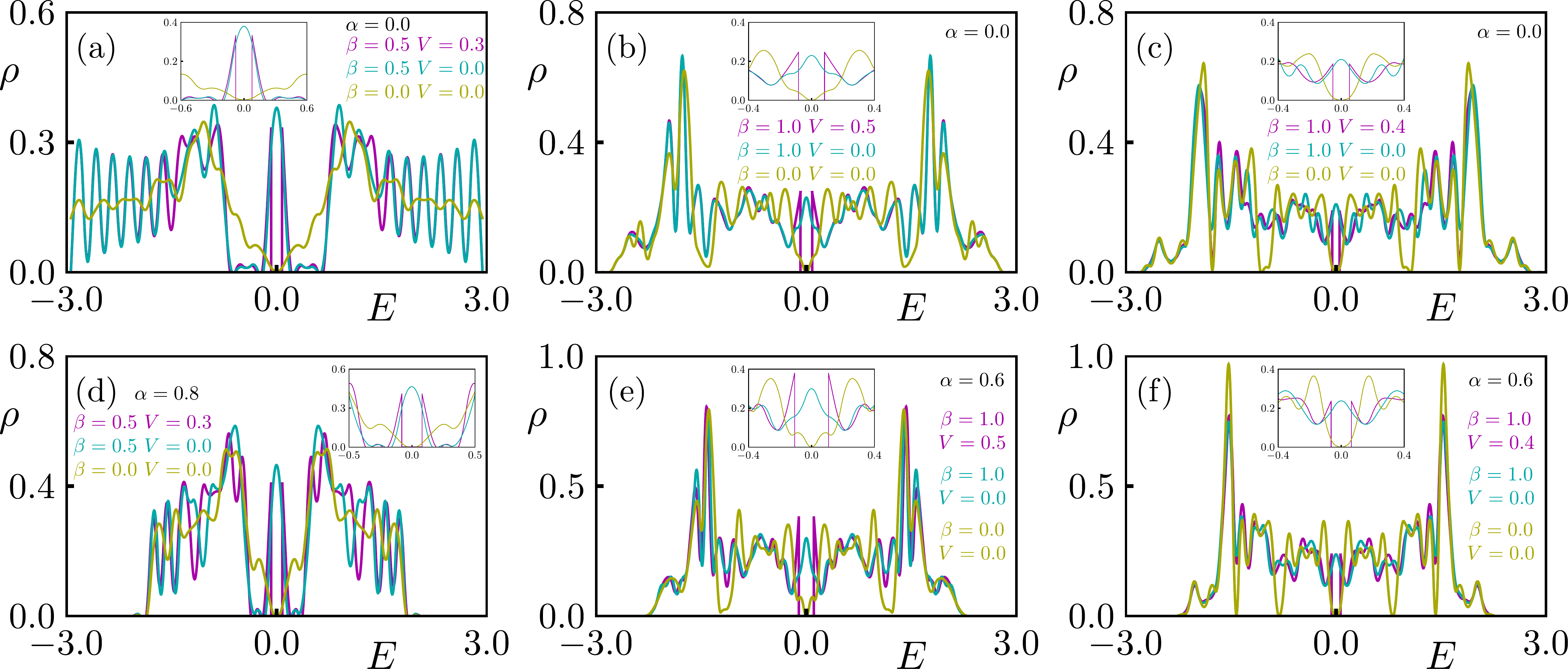}
    \caption{Density of states (DOS) $\rho$ as a function of energy in Hermitian ($\alpha=0$) (a) Euclidean $\{ 6,3 \}$ or honeycomb, (b) hyperbolic $\{10, 3\}$, and (c) hyperbolic $\{14, 3\}$ lattices for zero ($\beta=0$) and finite uniform magnetic flux (finite $\beta$) through each plaquette in the absence ($V=0$) and for subcritical strength (finite $V$) of nearest-neighbor Coulomb repulsion (see Table~\ref{tab:citicalvalues}). See legends for specifications of the parameter values. The insets display zoomed-in pictures of the DOS near the zero-energy. With no magnetic field ($\beta=0$) and interaction ($V=0$), the DOS in all these systems vanishes near zero energy, consistent with their Dirac liquid classification (yellow curves). Introduction of a finite and strong uniform magnetic field in non-interacting systems yields a nonzero DOS near zero energy, as predicted by index theorems (cyan curves). Finally, upon turning on subcritical (with respect to the zero magnetic field case) interactions (finite $V$), a gap forms in the all these systems (magenta curves), thereby confirming that the magnetic field catalyzes spontaneous breaking of the sublattice symmetry through the formation of a charge-density-wave order for subcritical $V$. Panels (d), (e), and (f) are analogous to (a), (b), and (c), respectively, but in non-Hermitian systems with a non-zero $\alpha$, showing that the magnetic catalysis mechanism is equally operative in non-Hermitian Euclidean and hyperbolic Dirac systems for a commuting class mass (charge-density-wave in this case). For detailed discussion, see Sec.~\ref{sec:coulomb}. The DOS in honeycomb lattice without magnetic fields and in the absence of interactions is obtained with periodic boundary conditions to suppress the imprints of topological boundary modes living on its zigzag edges.
    }~\label{fig:3_DOSCDW}
\end{figure*}

On the honeycomb lattice, fostering massless Dirac fermions living on a flat Euclidean plane, we observe a clear formation of the zeroth Landau level at $E=0$ and a few additional particle-hole symmetric discrete Landau levels at finite and well-spaced energies where the DOS shows prominent spikes, when the system is subject to uniform magnetic fields in both Hermitian ($\alpha=0$) and NH (finite $\alpha$) setups. The number of states within the zeroth Landau level does not change with increasing $\alpha$, which is primarily confirmed from the independence of the spike height for the DOS of the zeroth Landau level for various $\alpha$, while the rest of the spectrum gets progressively squeezed toward the band center or zero-energy as $\alpha$ increases. These outcomes are shown in Fig.~\ref{fig:2_DOSFixedBVaryA}(a). We also note that the DOS near the zero-energy on honeycomb lattice scales linearly with $\beta$ or the total flux enclosed by the system, in agreement with the predictions from the Aharonov-Casher index theorem~\cite{aharonovcasher}, which remains operative on honeycomb lattice~\cite{HF:1, krishanuroy}. Furthermore, with increasing $\alpha$ the DOS near zero-energy does not change in the system, which is also in agreement with the same index theorem, which has recently been extended to Lorentz invariant NH Dirac system, living on the flat Euclidean space~\cite{NHAharonovCashed}. These findings are shown in Fig.~\ref{fig:2_DOSFixedBVaryA}(d).

On $\{ 10,3 \}$ and $\{ 14, 3 \}$ lattices, accommodating massless Dirac fermions in the hyperbolic plane, application of even uniform magnetic fields does not produce any Landau levels at finite energies due to the constant negative spatial curvature~\cite{Bfieldhyperbolic:0, Bfieldhyperbolic:1, Bfieldhyperbolic:2}. Nonetheless, the DOS near zero-energy then becomes finite~\cite{Bfieldhyperbolic:3}. We believe that such observation stems from a generalized Aharonov-Casher index theorem on hyperbolic space, although exact analytical proof of this statement is still missing. The rest of the spectrum at finite energies remains continuous and there is no Landau gap therein. While the zero-energy modes do not get affected by the introduction of non-Hermiticity in the system, the rest of the spectrum gets increasingly squeezed toward zero-energy with larger $\alpha$. As a result, when we compute DOS near the zero-energy in the presence of a continuous spectrum occupying the region with finite $E$, it shows a small increment with increasing $\alpha$. These numerically observed features are shown in Figs.~\ref{fig:2_DOSFixedBVaryA}(b) and~\ref{fig:2_DOSFixedBVaryA}(c) for $\{ 10,3 \}$ and $\{ 14, 3 \}$ hyperbolic lattices, respectively.

The DOS near the zero-energy for hyperbolic Dirac systems subject to uniform magnetic fields scales almost quadratically with the magnetic field strength or the total magnetic flux enclosed by the system. Although such a feature is observed in both Hermitian and NH systems, the quality of the quadratic fit obtained from our numerical analysis gets slightly worse with increasing $\alpha$ due to the continuous nature of the energy spectrum around the zero energy for any $\alpha$ (absence of any Landau gap) and the increase in the number of states around zero-energy with increasing $\alpha$, despite the number of exact zero-energy modes being insensitive to $\alpha$. These observations are summarized in Figs.~\ref{fig:2_DOSFixedBVaryA}(e) and~\ref{fig:2_DOSFixedBVaryA}(f) for $\{ 10,3 \}$ and $\{ 14, 3 \}$ hyperbolic lattices, respectively. With the existence of a large (finite) DOS near the zero-energy on Euclidean (hyperbolic) Dirac systems being unambiguously established in Hermitian and NH systems, next we proceed to unfold the effects of electronic interaction in such systems, when they are at half-filling, with the aim to promote the NH amplification of magnetic catalysis mechanism for commuting class masses (defined below).

\section{Magnetic catalysis and its Non-Hermitian amplification}~\label{sec:NHamplMagCata}

In this section, we discuss the instability of such NH Dirac systems, subject to external strong magnetic fields that produces large (finite) DOS near the zero-energy, when embedded on a Euclidean (hyperbolic) plane. Notice that the existence of such a robust zero-energy manifold allows the formation of various broken symmetry phases even for infinitesimally weak interactions. Among candidate ordered phases, the mass orders that are represented by matrix operators ($\hat{{\mathcal O}}$) that fully anticommute with the free-fermion NH operator $\hat{h}_{\rm NH}$ is energetically most favored as they not only gain energy by splitting the zero-energy manifold, but also push all the filled states at negative energies further down. In the context of Hermitian Dirac systems, the formation of mass orders through this mechanism is called \emph{magnetic catalysis}. Here, we set out to establish the jurisdiction of this mechanism in NH Dirac systems. Notice that for $\hat{{\mathcal O}}$ to fully anticommute with $\hat{h}_{\rm NH}$, such that it qualifies as a mass order in NH Dirac systems, two (anti)commutation conditions have to be satisfied simultaneously; (1) $\{ \hat{h}_0, \hat{{\mathcal O}} \}=0$ (to qualify as a mass in Hermitian system) and (2) $[\hat{h}_{\rm mass}, \hat{{\mathcal O}}]=0$ such that $\hat{{\mathcal O}}$ anticommutes with the anti-Hermitian component of $\hat{h}_{\rm NH}$. It is because of the second condition, we name such mass orders that can maximally gain condensation energy as \emph{commuting class masses}~\cite{NHcatalysis:1, NHDirac:1}.

With the specific construction of $\hat{h}_{\rm NH}$ or choice of $\hat{h}_{\rm mass}$ [Eq.~\eqref{eq:hmass}], we immediately identify two candidates belonging to the commuting class mass family, which are represented by the matrix operators 
\begin{eqnarray}~\label{eq:massoperators}
\hat{\mathcal O} \equiv \hat{\mathcal O}_{\rm CDW} &=& \sigma_0 \otimes \left( 
\begin{array}{cc}
{\boldsymbol \Delta} & {\boldsymbol 0} \\
{\boldsymbol 0} & -{\boldsymbol \Delta}
\end{array}
\right) \nonumber \\
\text{and} \: \hat{\mathcal O} \equiv \hat{\mathcal O}_{\rm AFM} &=& \sigma_j \otimes \left( 
\begin{array}{cc}
{\boldsymbol \Delta} & {\boldsymbol 0} \\
{\boldsymbol 0} & -{\boldsymbol \Delta}
\end{array}
\right),
\end{eqnarray} 
where $j\in\{1,2,3\}$, and $\boldsymbol{\Delta}$ is an $N/2$-dimensional diagonal matrix, whose entries can be the same or different. Respectively, these two operators represent the CDW and AFM orders, displaying staggered patterns of electronic density and spin between two sublattices $A$ and $B$.

Next, we put forward an argument in favor of the magnetic catalysis of commuting class masses that gets amplified by the non-Hermiticity in Euclidean and hyperbolic Dirac systems, subject to external strong perpendicular magnetic fields. Since any commuting class mass by definition anticommutes with the total NH operator $\hat{h}_{\rm NH}$, we can immediately arrive at the following conclusion regarding the eigenspectrum of the effective single particle operator in the presence of such a mass order. For the sake of simplicity, we consider a uniform amplitude $\Delta$ of a commuting class mass order in the entire system. Then the zero-energy manifold of $\hat{h}_{\rm NH}$ gets split into two sets of states at energies $\pm E^\Delta_0$, where $E^\Delta_0 =\Delta$, and any finite energy state (at positive or negative energy) of $\hat{h}_{\rm NH}$ at energy $\sqrt{1-\alpha^2} E_i$ gets pushed to $ E^\Delta_i$, where $E^\Delta_i=\sqrt{(1-\alpha^2) E^2_i + \Delta^2}$ and $E_i$ depends on the magnetic field strength $\beta$. For generality, we assume the states at $\pm E^\Delta_i$ have a $D_i$-fold degeneracy per unit area or per site for any $i=0, 1, \cdots$. With these assumptions, we arrive at the following expression for the free-energy per site at zero temperature at half-filling
\begin{eqnarray}
f = \frac{\Delta^2}{2g} - \sum_{i=0,1, \cdots} D_i \; E^\Delta_i 
\end{eqnarray}
where $g$ is the four-fermion coupling constant favoring the commuting class mass order. It should be noted that the degeneracy of any state is independent of $\alpha$.

Minimizing $f$ with respect to $\Delta$, we arrive at the self-consistent mean-field gap equation which besides a trivial solution of $\Delta=0$ also supports a non-trivial solution for $\Delta$, which is obtained by solving 
\begin{equation}
\frac{1}{g} = \frac{D_0}{\Delta} + \sum_{i=1,2,\cdots} \: \frac{D_i}{\sqrt{(1-\alpha^2) E^2_i + \Delta^2}}. 
\end{equation}
Although without the explicit notion of all the eigenvalues $E_i$, we cannot find the exact solution of $\Delta$ as a function of $g$ and $\beta$, the above gap equation is sufficient to first establish the magnetic catalysis in Hermitian systems and subsequently its amplification due to the non-Hermiticity in the system.

Notice that for any fixed $\alpha$ (including the Hermitian case with $\alpha=0$), a finite solution of $\Delta$ can be found for any coupling constant $g$. It can be appreciated by neglecting the second term in the above gap equation, yielding then $g \sim \Delta$, implying the magnetic catalysis that takes place due to a large or finite DOS near zero-energy. When the positive-definite contribution from the entire filled Dirac sea (second term) is taken into account, the self-consistent solution for $\Delta$ for a fixed coupling constant $g$ becomes bigger in comparison to that when the contribution of the filled Dirac sea is completely neglected. This is a crucial observation which plays an important role behind the magnetic catalysis of Dirac masses. Namely, nucleation of Dirac masses not only gains energy by splitting the zero-energy manifold, it gains additional condensation energy by pushing the entire filled Dirac sea further down in energy.

Finally, we scrutinize the impacts of $\alpha$ on the self consistent solution of $\Delta$ from the above gap equation. For concreteness, we maintain a fixed value of the coupling constant $g$ and magnetic field strength $\beta$ for the rest of the discussion in this paragraph. First notice that the contribution of the zero-energy manifold (first term in the gap equation) is insensitive to the non-Hermiticity ($\alpha$) in the system. Nonetheless, with increasing $\alpha$, as the contribution from the term $(1-\alpha^2) E^2_i$ gets weaker (accounting for the contribution of the filled Dirac sea), the self-consistent value for $\Delta$ must get bigger when $g$ is held fixed. This outcome holds for any arbitrary $\beta$. In other words, with increasing $\alpha$, the magnitude of the commuting class mass amplifies for a fixed strength of the coupling constant and magnetic field, which is the statement of our proposed non-Hermitian amplification of the magnetic catalysis mechanism. Next, we anchor this proposal from lattice-based numerical self-consistent solutions for CDW and AFM orders, two prominent members of the commuting class mass family, in our specific construction of NH Dirac systems with a particular choice of $\hat{h}_{\rm mass}$ [see Eq.~\eqref{eq:hmass}].

\section{Biorthogonal Quantum Mechanics}~\label{sec:biorthogonal}

Since we are working with NH models, a review on some essential concepts from the biorthogonal quantum mechanics that we will be using in this work are worth reviewing briefly~\cite{biorthogonalQM}. Let us denote a generic NH operator as  $\mathcal{H}$ with $\mathcal{H} \neq \mathcal{H}^{\dagger}$. The left ($L$) and right ($R$) eigenvectors of ${\mathcal H}$ with eigenvalue $E$ (generically complex) are obtained from the eigenvalue equations 
\begin{equation}
    \mathcal{H} \ket{E^{\rm R}} = E\ket{E^{\rm R}}\; \text{and}\; \bra{E^{\rm L}} \mathcal{H} = \bra{E^{\rm L}} E^\star,
\end{equation}
respectively, where `$\star$' denotes the complex conjugation. However, in this work we deal with only such NH operators that feature all-real eigenvalue spectra for which $E=E^\star$. Unlike in the case of Hermitian quantum mechanics, the right and left eigenvectors are no longer related to one another by a simple complex conjugation transpose operation $(\bra{E_{\rm L}})^{\dagger} \neq \ket{E_{\rm R}}$. Accordingly, the orthonormality condition is ensured in terms of the biorthogonal product according to
\begin{equation}~\label{eq:birothogonal}
    \bra{E_{a}^{\rm L}} \ket{E_{b}^{\rm R}} = \delta_{a,b},
\end{equation}
where $\delta_{a,b}$ is the Kronecker delta symbol, and $a$ and $b$ index the eigenvalues. Extending the biorthogonal condition, we find that the probability of finding a particle with an eigenvalue $E_i$ at position $j$ is given by
\begin{equation}
    P = \bra{E_a^{\rm L}}\ket{j^{\rm R}} \bra{j^{\rm L}}\ket{E_a^{\rm R}}
\end{equation}
where $\bra{j^{\rm L}}$ ($\ket{j^{\rm R}}$) is the site-localized Wannier left (right) state on the $j$th site. We use this definition of $P$ while computing the CDW order self-consistently within the mean-field approximation in Sec.~\ref{sec:coulomb}. This definition can immediately be generalized to compute the average occupation number for each spin projection on each site separately (not shown here explicitly), which we use to compute the AFM order self-consistently, see Sec.~\ref{sec:hubbard}.

Following from the biorthogonal condition given in Eq.~\eqref{eq:birothogonal}, left and right eigenvectors are obtained numerically and normalized accordingly. However, while numerical eigensolvers specialized for Hermitian operators are robust due to their ability to leverage the guaranteed set of resulting orthonormal eigenvectors, generalized eigensolvers (employed here for NH operators) do not always necessarily enjoy such a stability. For example, when a generalized eigensolver is provided with an operator whose spectrum is non-degenerate, but has eigenvalues residing close to one another, numerical instabilities can result in eigenvectors that violate the biorthogonal condition from Eq.~\eqref{eq:birothogonal}. To stabilize these generalized eigensolvers, a small amount of on-site disorder is added to the system to restore the biorthogonality condition. The on-site disorder is drawn from a uniform box distribution $[-W/2, W/2]$, where $W$ denotes the disorder strength and we typically set $W \sim 10^{-3}$. Such a small amount of disorder causes only a negligible shift of Fermi energy.

\begin{figure*}[t!]
    \includegraphics[width=\linewidth]{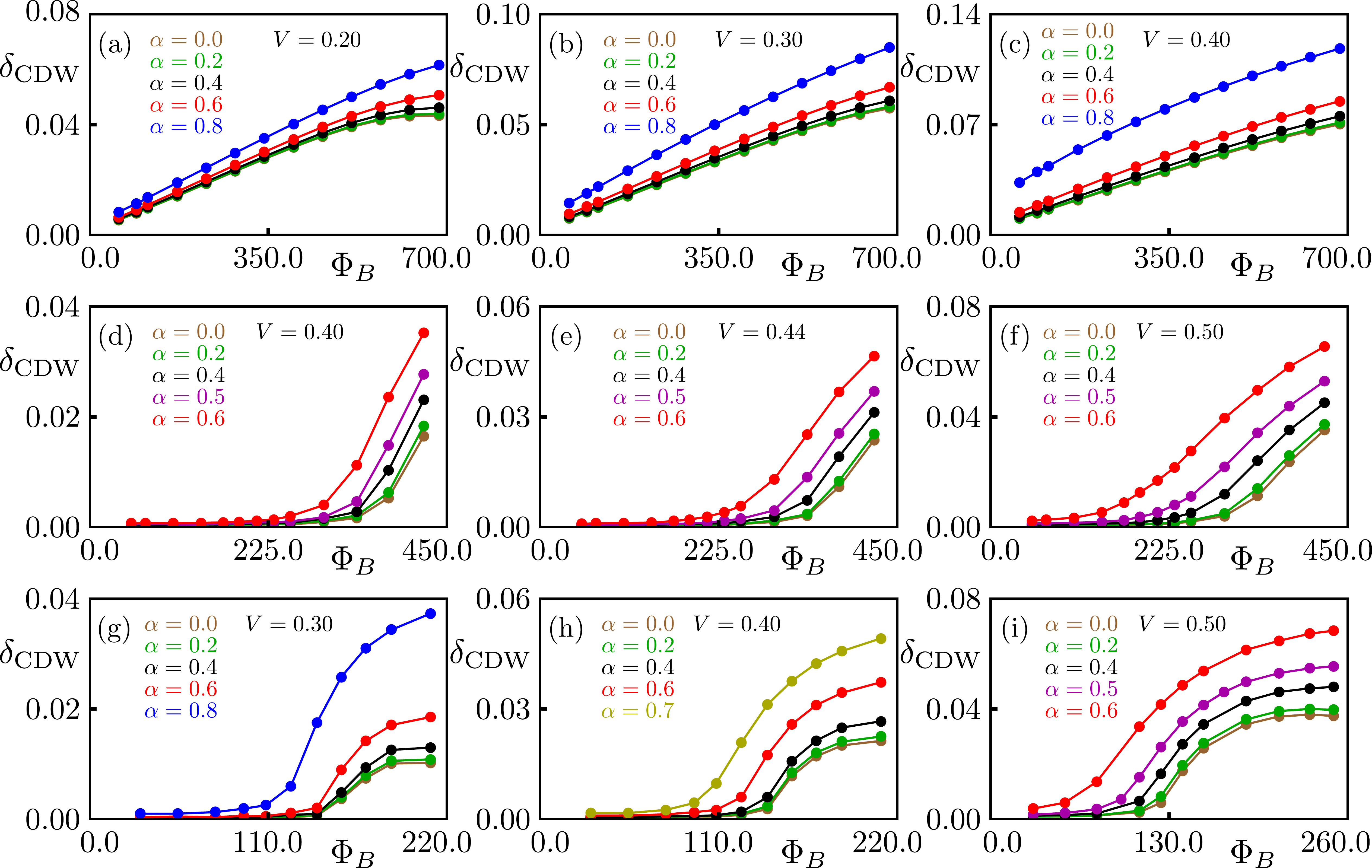}
    \caption{The scaling of the charge-density-wave (CDW) order parameter $\delta_{\rm CDW}$ as a function of the total magnetic flux $\Phi_B$ with varying degree of non-Hermiticity $\alpha$ and a few fixed (in a given subfigure) strengths of subcritical nearest-neighbor Coulomb repulsion $V$ in Euclidean $\{6, 3\}$ or honeycomb [(a)-(c)], hyperbolic $\{10, 3\}$ [(d)-(f)], and hyperbolic $\{14, 3\}$ [(g)-(i)] lattices. In all cases, it is evident that with increasing total magnetic flux enclosed by the system ($\Phi_B$), the amplitude of the CDW order increases. The emergence of the CDW order at subcritical strengths of nearest-neighbor Coulomb repulsion (see Table~\ref{tab:citicalvalues}) endorses the magnetic catalysis in these systems, regardless of the underlying spatial curvature and the degree of non-Hermiticity in the system. Most importantly, with increasing non-Hermiticity (larger $\alpha$) magnitude of the CDW order amplifies monotonically when $\beta$ and $V$ are held fixed, thereby promoting the proposed mechanism of amplified magnetic catalysis of commuting class masses (CDW in this case). See legends for specific parameter values and Sec.~\ref{sec:coulomb} for a detailed discussion.  
    }~\label{fig:5_CDWOrderVaryB}
\end{figure*}

\section{NN Coulomb Repulsions and CDW}~\label{sec:coulomb}

Motivated by the analytic predictions on the NH amplification of magnetic catalysis of commuting class masses, we, in this section, set out to anchor them from the numerical self-consistent solutions for the CDW order, a member of the commuting class mass family with the specific realization of a NH Dirac system discussed in Sec.~\ref{sec:nhfree}. The CDW order, featuring a staggered pattern of electronic density between the sites of two sublattices of any bipartite lattice, emerges as a natural ground state of NN Coulomb interaction in half-filled systems~\cite{NN:1, NN:2, NN:3}, which can be shown by decomposing the corresponding four-fermion term in the Hartree channel. Furthermore, condensation of the CDW order gets naturally catalyzed by the presence of external strong magnetic fields giving birth to a topologically robust near zero-energy manifold in Euclidean and hyperbolic Dirac systems. Owing to these reasons, here we include the NN Coulomb repulsion into our study, which then becomes a prototypical model to stage the formation of a mass order that can be amplified by the NH degree of freedom in the system.

The Hamiltonian for NN Coulomb repulsion reads as 
\begin{equation}~\label{eq:nncoulomb}
    H_{V} = \frac{V}{2} \; \sum\limits_{ \langle i, j \rangle} \; n_{i} n_{j} - \mu N,
\end{equation}
where $V$ is a real-valued parameter describing the coupling strength of the interaction, $\langle \cdots \rangle$ restricts the summation to NN sites, $n_{i}$ ($n_{j}$) is the number density operator on the $i$th ($j$th) site, $\mu$ is the chemical potential measured from zero-energy or half-filling, and $N$ is the total number of sites in the system. Since the CDW order is insensitive to the spin degrees of freedom [see Eq.~\eqref{eq:massoperators}], here for the sake of simplicity, we consider a collection of spinless fermions. After decomposing the four-fermion term from Eq.~\eqref{eq:nncoulomb} in the Hartree channel, we arrive at the following effective single-particle Hamiltonian~\cite{HF:1, HF:2, HF:3, HF:4} 
\begin{equation}
    H_V^{\rm Har} = V\sum\limits_{ \langle i, j \rangle} \left ( \langle n_{A,i} \rangle \; n_{B,j} + \langle n_{B,i} \rangle \;  n_{A,j} \right ) - \mu N
    \label{eq:coulomb_mft}
\end{equation}
where $n_{A,i}$ ($n_{B,j}$) and $\langle n_{A,i} \rangle$ ($\langle n_{B,j} \rangle$) are the number operator and average fermionic density for the $i$th ($j$th) site belonging the $A$ ($B$) sublattice, respectively, and, to maintain the half-filling, $\mu$ is set to $V/2$. Upon adding the matrix operator associated with Eq.~\eqref{eq:coulomb_mft}, say $\hat{h}^{\rm Har}_{\rm Coul}$ to the free-fermion operator $\hat{h}_{\rm NH}$, capturing the imprints of external magnetic fields and non-Hermiticity, we self-consistently compute the site-dependent quantities $\langle n_{A,i} \rangle=1/2 + \delta_{A,i}$ and $\langle n_{B,j} \rangle=1/2 -\delta_{B,j}$ with $\delta_{A/B,j}>0$, both being measured from their values in the half-filled non-interacting systems, equal to $1/2$, following the biorthogonal definition of probability density (Sec.~\ref{sec:biorthogonal}). Without any loss of generality, we assume that the average electronic density on the sites of the $A$ ($B$) sublattice increases (decreases) with respect to the uniform background value in the non-interacting half-filled system due to the spontaneous breaking of the Ising-like sublattice exchange symmetry.

We begin by scrutinizing the reconstructed DOS from the self consistent solutions of the CDW order, in terms of $\delta_{A,i}$ and $\delta_{B,j}$ for subcritical strengths of the NN Coulomb repulsions in the Euclidean $\{ 6,3\}$ or honeycomb lattice and the hyperbolic $\{10, 3\}$ and $\{14, 3\}$ lattices in Hermitian ($\alpha=0$) systems and in the presence of non-Hermiticity (finite $\alpha$) in the system. The results of which are shown in Fig.~\ref{fig:3_DOSCDW}. Firstly, we note that all these systems display the characteristic linearly vanishing DOS near zero-energy in the absence of any external magnetic field and interaction, irrespective of the non-Hermiticity therein. In the presence of external magnetic fields, the non-interacting fermions on these Euclidean and hyperbolic lattices display a large and a finite DOS near the zero-energy, respectively. This conclusion holds both in Hermitian and non-Hermitian systems, which is consistent with the predictions from the index theorems. Formation of such non-zero DOS near the zero-energy becomes conducive to the nucleation of the CDW order in Hermitian systems even with subcritical strengths of $V$, a phenomenon we know by the name of magnetic catalysis. This outcome can be confirmed from the DOS near the zero-energy for subcritical $V$ for which a clear gap is observed therein. In NH systems, the CDW order represents a member of the commuting class mass family and thus can be supported by external strong magnetic fields for subcritical $V$, as discussed in the previous section. This prediction is once again confirmed from the sharp gap in DOS near the zero-energy in all the NH Dirac systems, placed in uniform strong magnetic fields for the corresponding subcritical strength of $V$, see Table~\ref{tab:citicalvalues}. Hence, magnetic catalysis for planar Euclidean and hyperbolic Dirac liquids remains operative for commuting class masses in NH setups. By comparing such a gap size in DOS near the zero-energy for various $\alpha$, we conclude that increasing non-Hermiticity amplifies this gap, endorsing the NH amplification of the magnetic catalysis mechanism, see Fig.~\ref{fig:Summary} (top row).

\begin{figure*}[t!]
    \includegraphics[width=\linewidth]{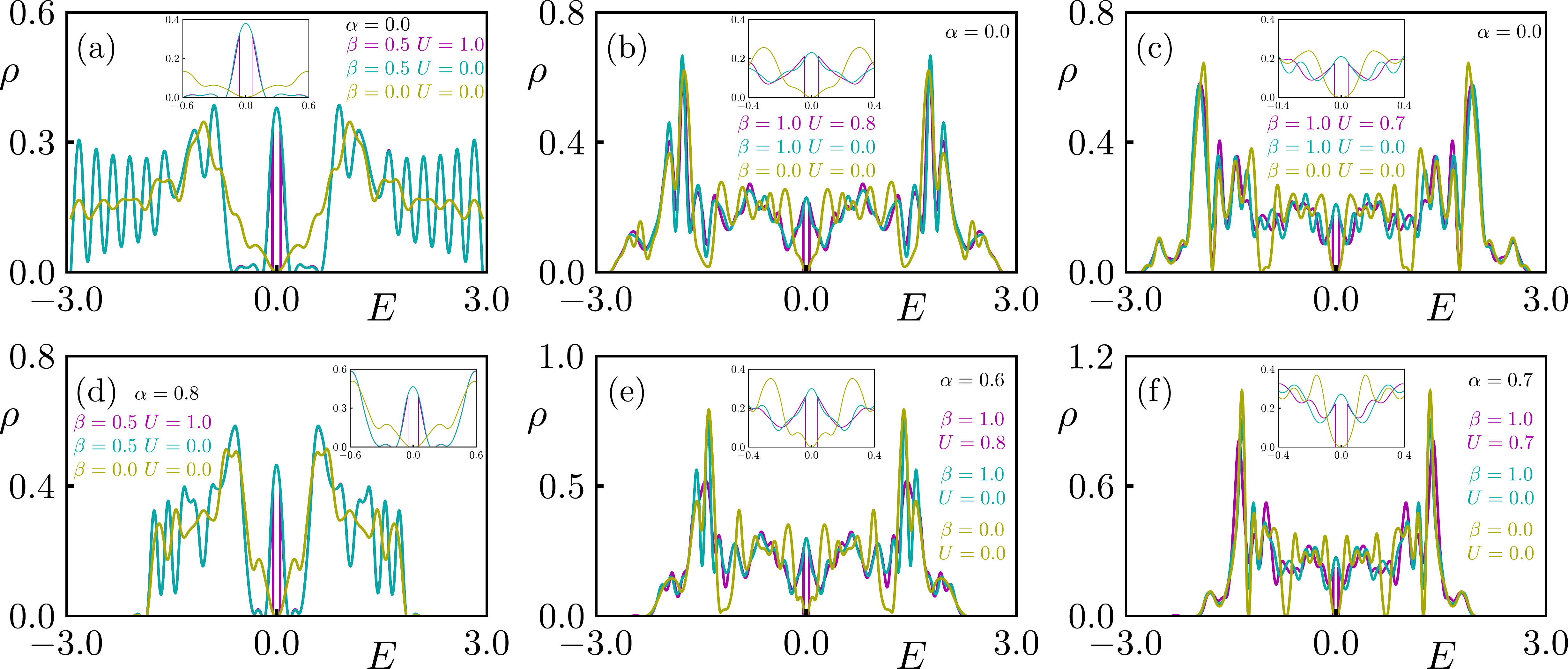}
    \caption{Analogous to Fig.~\ref{fig:3_DOSCDW}, but density of states (DOS) in interacting systems are computed in the presence of subcritical on-site Hubbard repulsion ($U$), see Table~\ref{tab:citicalvalues}. DOS of Hermitian ($\alpha=0$) (a) Euclidean $\{ 6,3\}$ or  honeycomb, (b) $\{10, 3\}$ hyperbolic, and (c) $\{14, 3\}$ hyperbolic lattices in non-interacting systems ($U=0$), without any magnetic field ($\beta=0$) and for finite $\beta$ are shown by yellow and cyan curves, respectively, displaying vanishing and finite DOS near zero-energy. Even for subcritical strength of on-site Hubbard repulsion, such systems become insulators at half-filling, see magenta curves, confirming the magnetic catalysis for antiferromagnetic Dirac mass. We arrive at similar conclusions with finite non-Hermiticity ($\alpha$) in the system, as shown for (d) Euclidean $\{ 6,3\}$ or  honeycomb, (e) $\{10, 3\}$ hyperbolic, and (f) $\{14, 3\}$ hyperbolic lattices. These findings establish magnetic catalysis for a commuting class mass (antiferromagnet in this case) in non-Hermitian Euclidean and hyperbolic Dirac systems. For detailed discussion, see Sec.~\ref{sec:hubbard}. See legends for specifications on various parameter values. The DOS in honeycomb lattice without magnetic fields and in the absence of interactions is obtained with periodic boundary conditions to suppress the imprints of topological boundary modes living on its zigzag edges.
    }~\label{fig:6_DOSAFM}
\end{figure*}

In order to further anchor and quantify the NH amplification of magnetic catalysis, operative over members of the commuting class mass family, next we study the scaling of the CDW order parameter over a wide range of uniform magnetic fields for a few choices of the NH parameter ($\alpha$) and subcritical strengths of NN Coulomb repulsion ($V$). For this analysis, we define the order parameter for the emergent CDW order after averaging it over the entire system, given by 
\begin{equation}~\label{eq:cdworderparam}
    \delta_{\rm CDW} = \frac{1}{N} \sum_{i} \left[ \left( \langle n_{A,i} \rangle -\frac{1}{2} \right) - \left( \langle n_{B,i} \rangle-\frac{1}{2} \right) \right],
\end{equation}
whose scaling behavior is shown in Fig.~\ref{fig:5_CDWOrderVaryB}. Irrespective of the amount of non-Hermiticity or the choice of subcritical interaction strength, both Euclidean and hyperbolic Dirac systems show an emergence of nonzero CDW order for strong magnetic fields, in qualitative agreement with our findings summarized in Fig. \ref{fig:3_DOSCDW} for a single choice of $V$. In a Euclidean Dirac system, for small subcritical $V$, the CDW order grows linearly and rapidly with the magnetic field due to the formation of sharp Landau levels, especially the zeroth one giving a large DOS near the zero-energy. By contrast, the growth of $\delta_{\rm CDW}$ with the magnetic field strength or total magnetic flux enclosed by the hyperbolic system is weak due to the lack of a sharp Landau quantization therein, yielding only a finite DOS near the zero-energy. These scaling features are observed for both Hermitian and non-Hermitian systems. But, the magnitude of $\delta_{\rm CDW}$ increases monotonically with $\alpha$ for any fixed values of $\beta$ and subcritical $V$. Altogether, this analysis further affirms the validity of the proposed scenario of NH amplification of magnetic catalysis for commuting class masses.

\begin{figure*}
    \includegraphics[width=\linewidth]{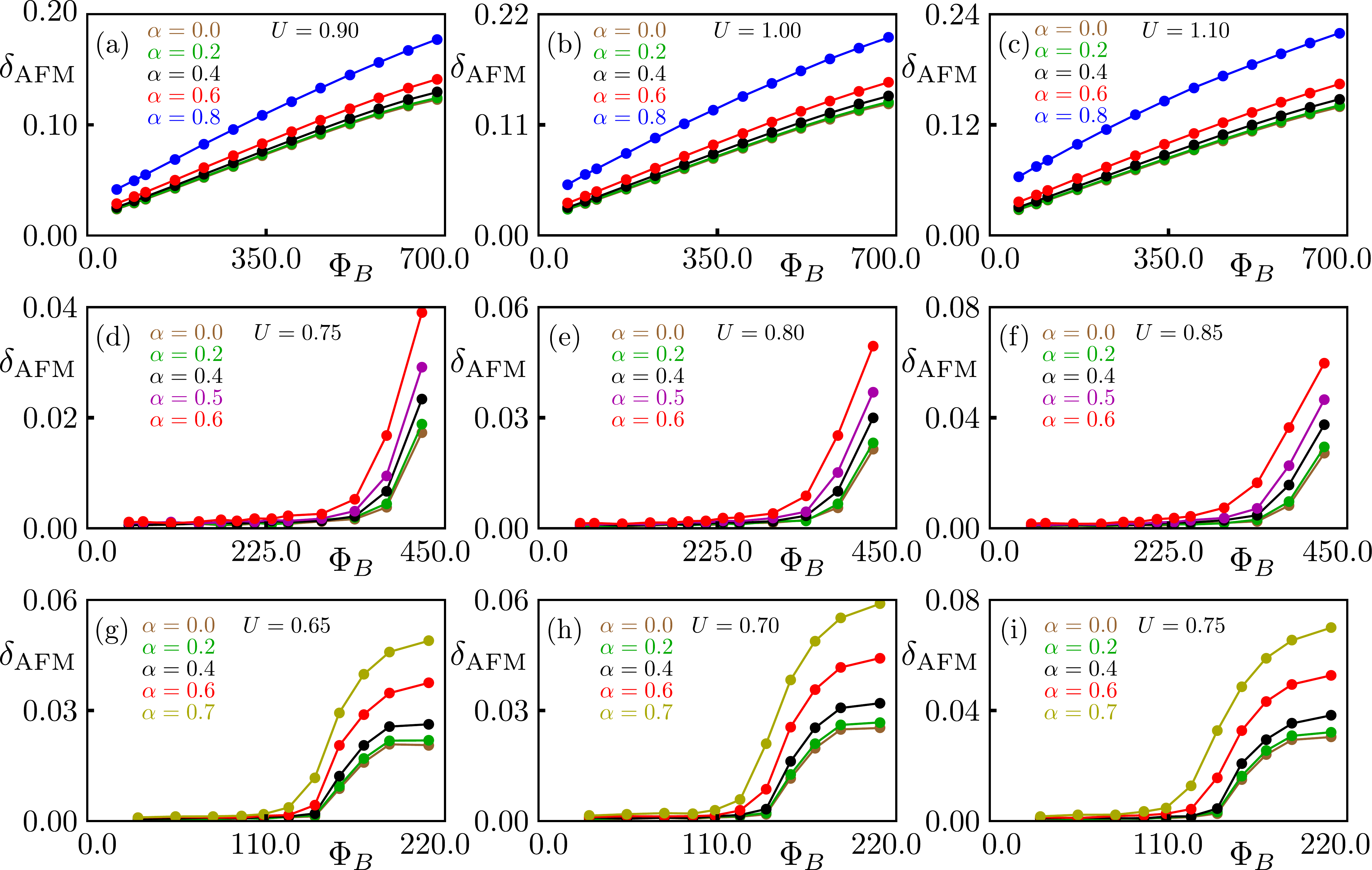}
    \caption{The scaling of the antiferromagnet (AFM) order parameter $\delta_{\rm AFM}$ as a function of the total magnetic flux $\Phi_B$ with varying degree of non-Hermiticity $\alpha$ and a few fixed (in a given subfigure) strengths of subcritical on-site Hubbard repulsion $U$ in Euclidean $\{6, 3\}$ or honeycomb [(a)-(c)], hyperbolic $\{10, 3\}$ [(d)-(f)], and hyperbolic $\{14, 3\}$ [(g)-(i)] lattices. In all cases, it is evident that with increasing total magnetic flux enclosed by the system ($\Phi_B$), the amplitude of the AFM order increases. The emergence of the AFM order at subcritical strengths of on-site Hubbard repulsion (see Table~\ref{tab:citicalvalues}) endorses the magnetic catalysis in these systems, regardless of the underlying spatial curvature and the degree of non-Hermiticity in the system. Most importantly, with increasing non-Hermiticity (larger $\alpha$) the magnitude of the AFM order amplifies monotonically when $\beta$ and $U$ are held fixed, thereby promoting the proposed mechanism of amplified magnetic catalysis of commuting class masses (AFM in this case). See legends for specifications of parameter values and Sec.~\ref{sec:hubbard} for detailed discussion. 
    }~\label{fig:8_AFMOrderVaryB}
\end{figure*}

\section{Hubbard repulsion and AFM}~\label{sec:hubbard}

The previous section was dedicated to demonstrating the NH amplification of magnetic catalysis in half-filled Euclidean and hyperbolic Dirac liquids with NN Coulomb repulsion, giving rise to sublattice exchange symmetry breaking CDW order, a member of the commuting class mass family. In order to demonstrate the universal aspect of such a mechanism, here we consider on-site Hubbard repulsion, which gives rise to an emergent staggered pattern of electronic spin between two sublattices of a bipartite lattice, known as the AFM order~\cite{hyperbolicclassification, HL:9, NN:1, onsite:1, onsite:2, onsite:3, onsite:4}, which is also a member of the commuting class mass family in our construction of NH Dirac models, see Sec.~\ref{sec:nhfree} and Eq.~\eqref{eq:NHGeneral}.

\begin{figure*}
    \includegraphics[width=1.0\linewidth]{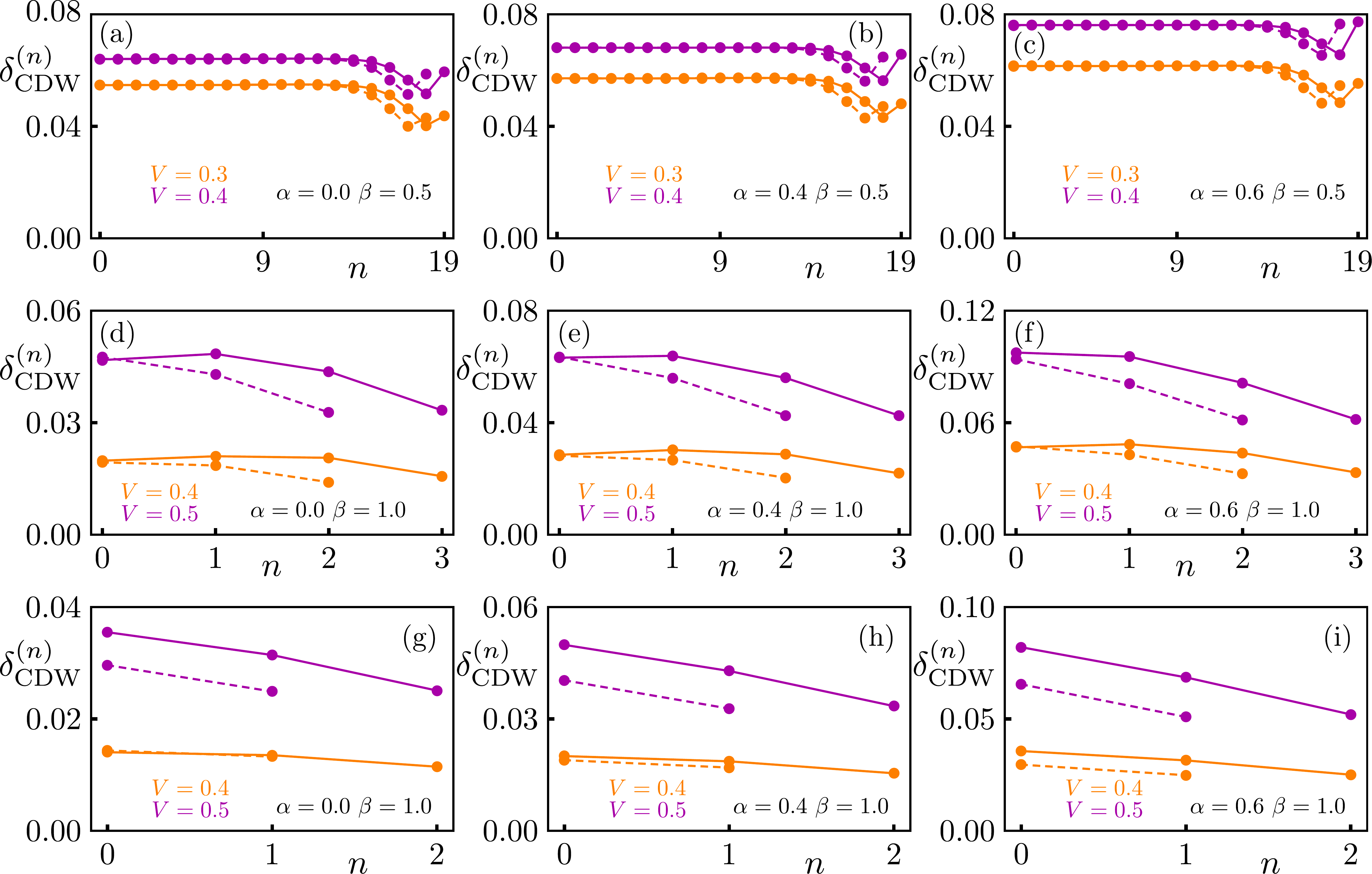}
    \caption{Finite size scaling analysis of the local charge-density-wave (CDW) order $\delta_{\rm CDW}^{(n)}$, defined over the sites of the $n$th generation (see Sec.~\ref{sec:system} for definition) of Euclidean $\{6, 3\}$ or honeycomb [(a-c)], hyperbolic $\{10, 3\}$ [(d-f)], and hyperbolic $\{14, 3\}$ [(g)-(i)] lattices with zero [(a), (d), and (g)], moderate [(b), (e), and (h)], and large [(c), (f), and (i)] non-Hermiticity $\alpha$, fixed nonzero magnetic flux per plaquette $\beta$, and two choices of the subcritical nearest-neighbor Coulomb repulsion $V$ (represented by color) in systems containing a total of $N$ (solid line) and $N-1$ (dashed line) number of generations with open boundary conditions. See also the legends of each subfigure for the parameter values. For all cases considered here, the CDW order in every generation of each system is nonzero, showing the emergence of CDW order throughout the entire system. Moreover, CDW orders in the bulk of a larger system with $N$ generation is approximately equal (on honeycomb lattice) or typically greater (on hyperbolic lattices) in magnitude than those in the bulk of a smaller system with $N-1$ generations. Also, the magnitude of the CDW order at the edges of these two systems, where it dips slightly when compared to its values in the bulk, are almost equal (within the numerical accuracy). In brief, these outcomes strongly suggest the robustness of the proposed amplified magnetic catalysis mechanism for commuting class masses in non-Hermitian Euclidean and hyperbolic Dirac systems (the CDW order in our construction) in the thermodynamics limit ($N \to \infty$). For a detailed discussion on the scaling of the local CDW order with the system size, $\alpha$, $V$, and $\beta$ see Sec.~\ref{sec:finitesize}. 
    }~\label{fig:9_finitesizeCDW}
\end{figure*}

The Hamiltonian for the Hubbard model is given by
\begin{equation}
    H_{\rm Hubb} = U \; \sum\limits_i \;  \left( n_{i, \uparrow} - \frac{1}{2} \right) \: \left( n_{i, \downarrow} - \frac{1}{2} \right) - \mu N,  
\end{equation}
where $U \; (>0)$ is a real-valued parameter describing the strength of the on-site repulsion, $n_{i, \uparrow}$ ($n_{i, \downarrow}$) is the number operator on the $i$th site with the electronic spin projection $\uparrow$ ($\downarrow$) in the $z$ direction, $\mu$ is the chemical potential measured from half-filling or zero-energy, and $N$ is the total number of sites in the system. We once again perform a Hartree decomposition to obtain an effective single-particle Hamiltonian, given by~\cite{HF:3, HF:4} 
\begin{equation}~\label{eq:hubbmft}
\begin{split}
    H_{\rm Hubb}^{\rm Har} &= U \; \sum\limits_i \; \bigg[ \left( \langle n_{i, \uparrow} \rangle - \frac{1}{2} \right) \left( n_{i, \downarrow} - \frac{1}{2} \right)\\
    &+ \left( \langle n_{i, \downarrow} \rangle - \frac{1}{2} \right) \left( n_{i, \uparrow} - \frac{1}{2} \right) \bigg] - \mu N,
\end{split}
\end{equation}
where $\langle n_{i, \uparrow} \rangle$ ($\langle n_{i, \downarrow} \rangle$) is the average electronic density on the $i$th site with spin projection $\uparrow$ ($\downarrow$). We work with the following ansatz  
\begin{equation}
\langle n_{A,\sigma} \rangle= \frac{1}{2} + \sigma \; \delta_{A,\sigma}(\vec{r}), 
\:\:\:
\langle n_{B,\sigma} \rangle= \frac{1}{2} - \sigma \; \delta_{B,\sigma}(\vec{r}) 
\end{equation}
suitable to capture AFM ordering for which $\delta_{A/B, \uparrow/\downarrow}>0$. The half-filling condition is maintained by choosing $\mu=0$ and
\begin{equation}
\sum_{\sigma=\pm} \; \sum_{\vec{r}} \; \sigma \; \delta_{A,\sigma} (\vec{r}) - 
\sum_{\sigma=\pm} \; \sum_{\vec{r}} \; \sigma \; \delta_{B, \sigma} (\vec{r}) =0,
\end{equation}
which also keeps the average electronic density equal to one at each site of the lattice (no charge ordering).

We add the operator associated with the effective single-particle particle Hamiltonian resulting from the Hartree decomposition of the on-site Hubbard model from Eq.~\eqref{eq:hubbmft}, say $\hat{h}^{\rm Har}_{\rm Hubb}$ to the free-fermion operator $\hat{h}_{\rm NH}$ with non-reciprocal complex hopping and compute $\delta_{A/B, \uparrow/\downarrow}$ self-consistently. From such solutions we compute the reconstructed DOS, shown in Fig.~\ref{fig:6_DOSAFM} and the findings are qualitatively similar to the ones we previously discussed with the CDW order (Fig.~\ref{fig:3_DOSCDW}). Namely, both Euclidean and hyperbolic non-interacting Dirac systems display linearly vanishing DOS in the absence of any magnetic field, while the DOS close to the zero-energy in these two types of Dirac systems become large and finite, respectively, in the presence of strong external magnetic fields, thereby creating a conducive platform for an insulation via the magnetic catalysis mechanism at arbitrarily weak interactions through the condensation of mass orders. This anticipation is confirmed from the observed finite gap even for subcritical on-site Hubbard repulsion near the zero-energy in Hermitian systems. Similar outcomes hold in NH Dirac systems, subject to external strong magnetic fields. Comparing the magnitude of the gap near the zero-energy for a fixed on-site repulsion, we find that it increases monotonically with increasing $\alpha$ in the system as long as $|\alpha|<1$, which can be verified explicitly from Fig.~\ref{fig:Summary} (bottom row), in turn confirming the proposed and predicted scenario of amplified magnetic catalysis of commuting class masses due to non-Hermiticity in Euclidean and hyperbolic Dirac systems.

\begin{figure*}
    \includegraphics[width=1.0\linewidth]{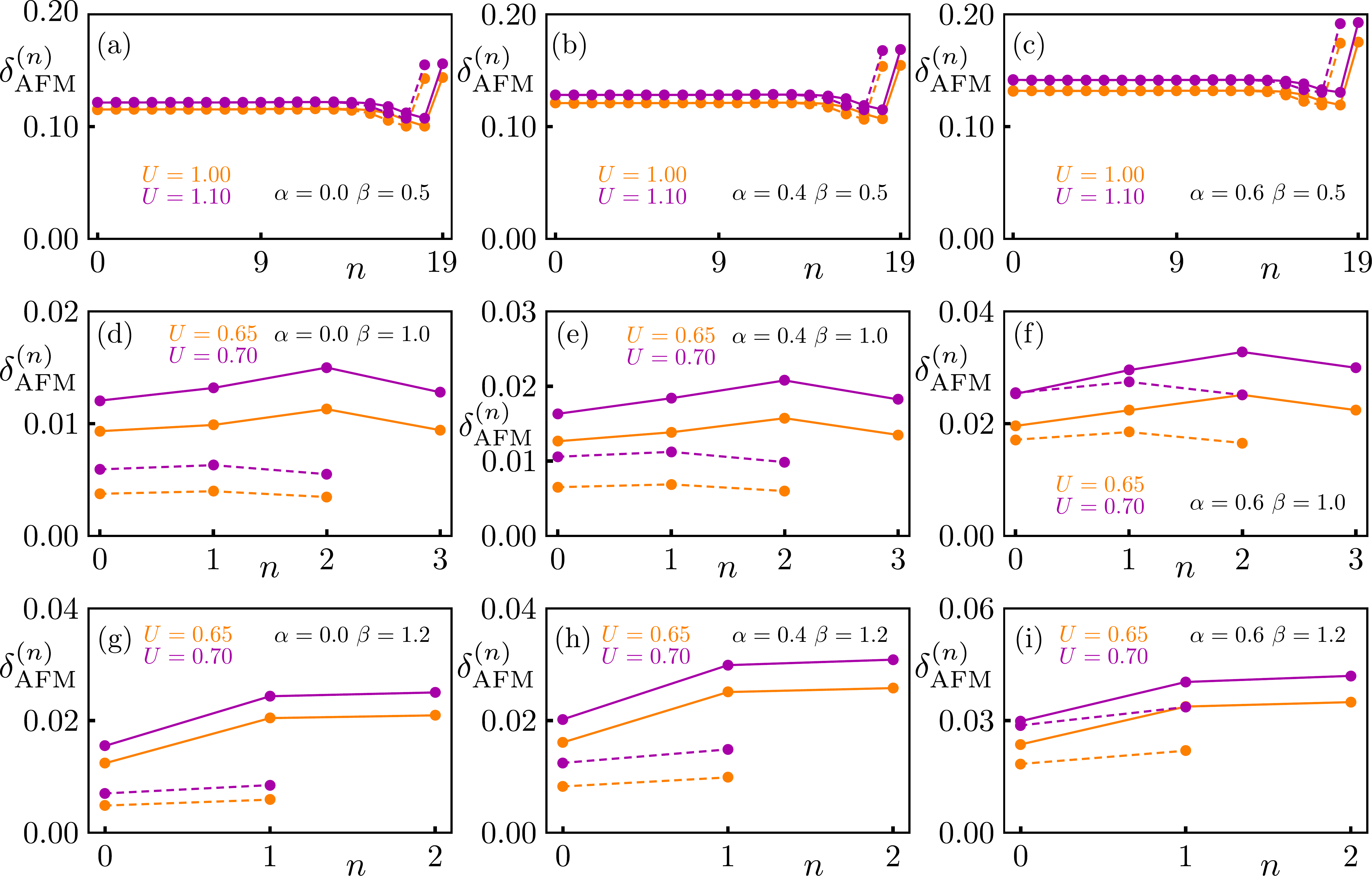}
    \caption{Finite size scaling analysis of the local antiferromagnet (AFM) order $\delta_{\rm AFM}^{(n)}$, defined over the sites of the $n$th generation (see Sec.~\ref{sec:system} for definition) of Euclidean $\{6, 3\}$ or honeycomb [(a-c)], hyperbolic $\{10, 3\}$ [(d-f)], and hyperbolic $\{14, 3\}$ [(g)-(i)] lattices with zero [(a), (d), and (g)], moderate [(b), (e), and (h)], and large [(c), (f), and (i)] non-Hermiticity $\alpha$, fixed nonzero magnetic flux per plaquette $\beta$, and two choices of the subcritical on-site Hubbard repulsion $U$ (represented by color) in two systems containing a total of $N$ (solid line) and $N-1$ (dashed line) number of generations with open boundary conditions. See also the legends of each subfigure for the parameter values. The rest of the details are the same as in the caption of Fig.~\ref{fig:9_finitesizeCDW} and for a detailed discussion on the scaling of the local AFM order with the system size, $\alpha$, $V$, and $\beta$ see Sec.~\ref{sec:finitesize}.
    }~\label{fig:10_finitesizeAFM}
\end{figure*}

Finally, we scrutinize the scaling of the AFM order over a wide range of external magnetic fields, and for a few choices of the non-Hermiticity and subcritical strength of the on-site repulsion in the system. For this purpose, we define the global AFM order parameter ($\delta_{\rm AFM}$) after averaging its local values over the entire system, given by  
\begin{equation}
\delta_{\rm AFM} = \frac{1}{N} \sum_{\vec{r}} \: \left[ \delta_{A, \uparrow} + \delta_{A, \downarrow} + \delta_{B, \uparrow} + \delta_{B, \downarrow} \right] (\vec{r}).
\end{equation}
The scaling behavior of $\delta_{\rm AFM}$ is shown in Fig.~\ref{fig:8_AFMOrderVaryB}, which is qualitatively similar to the one we previously discussed for the global CDW order $\delta_{\rm CDW}$, shown in Fig.~\ref{fig:5_CDWOrderVaryB} and thus we do not repeat it here. Most importantly, results shown in Fig.~\ref{fig:8_AFMOrderVaryB} unambiguously endorse the proposed mechanism of NH amplification of magnetic catalysis of commuting class masses in weakly interacting NH Dirac systems, embedded on a flat Euclidean or a constant negatively curved hyperbolic plane. Next, we set out to analyze the finite-size effects on CDW and AFM orders in these systems with open boundary conditions to establish the robustness of our predictions in the thermodynamic limit.

\section{Finite Size Analysis}~\label{sec:finitesize}

So far we have demonstrated the amplified magnetic catalysis of commuting class masses with a specific realization of NH Euclidean and hyperbolic Dirac systems from the self-consistent mean-field solutions of the CDW and AFM orders, favored by subcritical NN Coulomb and on-site Hubbard repulsions, respectively. This analysis has, however, been performed on the corresponding lattices with open boundary conditions, raising a natural question of whether such a mechanism holds in the thermodynamic limit. This section is dedicated to addressing this question, by studying the scaling of the local CDW ($\delta_{\rm CDW}^{(n)}$) and AFM ($\delta_{\rm AFM}^{(n)}$) order parameters, defined on the sites of each generation ($n$) of lattices with a total of $N-1$ and $N$ generations. See Sec.~\ref{sec:system} for the definition of the generation number. Specifically, we consider Euclidean honeycomb lattices with 19 and 20 generations, containing $2166$ and $2400$ sites, respectively, and $\{10, 3\}$ [$\{14, 3\}$] hyperbolic lattices with 2 and 3 [1 and 2] generations, containing $490$ and $2880$ sites [$168$ and $1694$ sites], respectively, each with a fixed strength of the uniform external magnetic field, but with a varying degree of non-Hermiticity in the system and for two subcritical interaction strengths.

We begin by considering such finite size effects on the local or generation-dependent CDW order in the presence of NN Coulomb repulsion. The analysis is shown in Fig.~\ref{fig:9_finitesizeCDW}. First, we observe that the CDW order in all generations of every system is nonzero, strongly indicating that the appearance of such an order with open boundary conditions is an emergent phenomenon, (possibly) occurring simultaneously throughout the entire system. Next, we compare the profile of $\delta_{\rm CDW}^{(n)}$ on a pair of systems with a total of $N-1$ and $N$ number of generations therein, allowing us to estimate the fate of the CDW order in the thermodynamic limit ($N \to \infty$). In all systems (see the previous paragraph and legends of Fig.~\ref{fig:9_finitesizeCDW} for explicit parameter values), the magnitude of the local CDW order in any fixed generation $n$ in a larger system with $N$ generations is approximately equal (within numerical accuracy) on honeycomb lattice or bigger (typically) on hyperbolic lattices than in a small system with $N-1$ generations for a fixed strength of the magnetic field, irrespective of the strength of the NN Coulomb repulsion ($V$) and non-Hermiticity ($\alpha$) in the system. Such a contrasting behavior of the local CDW order on honeycomb and hyperbolic lattices stems from the fact that with increasing total generation number, the total number of plaquettes in the system or the total flux enclosed by the system increases slowly and rapidly, respectively. This is a result of magnetic catalysis, since adding a generation with a fixed external uniform magnetic field results in more magnetic flux enclosed by the system, thereby giving rise to a stronger magnetic catalysis effect, which we have previously witnessed from the scaling of the global CDW order parameter [see Eq.~\eqref{eq:cdworderparam}], displayed in Fig.~\ref{fig:5_CDWOrderVaryB}. Finally, we note that the local CDW order parameter shows a dip near the edge of the system. However, we find that the magnitude of $\delta_{\rm CDW}^{(n)}$ at the edge of the system is almost equal (within numerical accuracy) on lattices with total of $N-1$ and $N$ generations for a fixed set of parameter values. This finding strongly suggests that edge values of the CDW order in systems with open boundary conditions are solely determined by the parameter values and independent of the system size. This observation is specifically important for hyperbolic lattices, where a large fraction of the sites belong to the one-dimensional edges of the system. Our findings, nonetheless, suggest that despite the presence of a large number of boundary sites the magnetic catalysis mechanism should remain operative on two-dimensional hyperbolic lattices in the thermodynamic limit.

Next, we focus on the impact of the NH parameter ($\alpha$) on these outcomes. We find that the overall variation of the local CDW order $\delta_{\rm CDW}^{(n)}$ remains unaffected by $\alpha$, but with increasing $\alpha$, $\delta_{\rm CDW}^{(n)}$ increases monotonically for any $n$, irrespective of the system size and strength of $V$ (subcritical). Finally, we note that the spatial variation of $\delta_{\rm CDW}^{(n)}$ is also insensitive to the subcritical strengths of $V$, while $\delta_{\rm CDW}^{(n)}$ increases with increasing $V$ for any $n$. The entire set of these observations is insensitive to the strength of the uniform magnetic magnetic fields, except that with increasing field strength $\delta_{\rm CDW}^{(n)}$ increases for any $n$ for any fixed values of the other tunable parameters (not shown in this work explicitly), which is expected in light of our findings, shown in Fig.~\ref{fig:5_CDWOrderVaryB}.

\begin{figure*}
    \includegraphics[width=\linewidth]{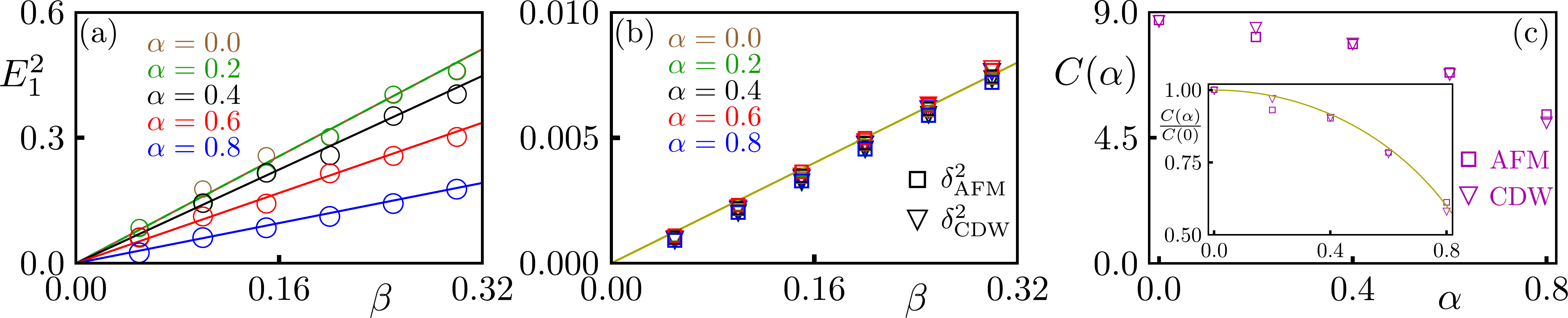}
    \caption{(a) Scaling of $E^2_1$, where $|E_1|$ is the first Landau level energy, with the magnetic field strength ($\beta$) in honeycomb lattice for different non-Hermiticity ($\alpha$) in the system that includes the Hermitian limit ($\alpha=0$). In all these cases we find linear fitting (within numerical accuracy), confirming the square-root dependence of the Landau level energy with the magnetic field in both Hermitian and non-Hermitian Euclidean Dirac systems. (b) Scaling of $\delta^2_{\rm CDW}$ and $\delta^2_{\rm AFM}$ (marked by different symbols) with $\beta$ for varying $\alpha$ (color-coded), where $\delta_{\rm CDW}$ ($\delta_{\rm AFM}$) is the self-consistent charge-density-wave (antiferromagnet) order parameter per spin projection, obtained for the corresponding $\alpha$-dependent critical strengths of the nearest-neighbor Coulomb ($V_c$) and on-site Hubbard ($U_c$) repulsions. See Table~\ref{tab:citicalvalues} for explicit values of $V_c$ and $U_c$ for all the choices of $\alpha$ on the honeycomb lattice. Within the numerical accuracy, we find that $\delta^2_{\rm CDW} \sim \beta$ and $\delta^2_{\rm AFM} \sim \beta$. (c) Scaling of the universal ratio $C(\alpha)=E_1/\delta_x$ for $x= {\rm CDW}$ and ${\rm AFM}$, with the non-Hermitian parameter, where isolated points are obtained from exact numerical calculations and the yellow line in the inset corresponds to the analytical predictions of $C(\alpha)/C(0)=\sqrt{1-\alpha^2}$. See Sec.~\ref{sec:summary} for discussions.      
    }~\label{fig:Honeyscaling}
\end{figure*}

Analogous analyses for the AFM order are shown in Fig.~\ref{fig:10_finitesizeAFM}. All the conclusions, we previously reported for the variation of the degeneration-dependent CDW order parameter, remain qualitatively the same for the local AFM order parameter $\delta_{\rm AFM}^{(n)}$, except for the following cases. The local AFM order shows a small upward spike near the boundary or edge of the honeycomb lattice in comparison to its values in the bulk or interior of the system. However, the magnitude of $\delta_{\rm AFM}^{(n)}$ at the edge of the system is insensitive to the system size and only depends on subcritical $U$, when all the other parameter values are kept unchanged (similar to what we find for the CDW order at the edge of the system). On hyperbolic lattices, the magnitude of the AFM order parameter at the edges is bigger in a larger system. Therefore, based on these observations we can conclude that magnetic catalysis of the CDW and AFM orders, members of the commuting class mass family with the specific realization of non-Hermiticity we considered in this work, takes place in NH Euclidean and hyperbolic Dirac systems, which gets amplified by the non-Hermiticity in the system, in the thermodynamic limit.

\section{Summary and discussions}~\label{sec:summary}

In this work, from extensive numerical analysis we unambiguously establish the amplified magnetic catalysis of commuting class masses triggered by the underlying non-Hermiticity in Dirac systems on Euclidean and hyperbolic lattices, when they are immersed in strong perpendicular magnetic fields. We propose a general robust argument in favor of this phenomenon, which we subsequently anchor from lattice-based numerical self-consistent solutions for the CDW and AFM orders in the mean-field limit after decomposing the NN Coulomb and on-site Hubbard repulsions in the Hartree channel, respectively. In our specific construction of lattice-regularized NH Dirac fermions on Euclidean $\{ 6,3 \}$ or honeycomb and hyperbolic $\{ 10,3 \}$ and $\{ 14,3 \}$ lattices, in which the non-Hermiticity results from an imbalance in the hopping amplitudes between the NN sites belonging to two sublattices in the opposite directions, both the CDW and AFM orders fall within the category of commuting class masses. While both the phases, being mass orders for Dirac fermions, get naturally nucleated at sufficiently weak interactions in the presence of external magnetic fields due to the formation a robust manifold of zero-energy modes, a mechanism named magnetic catalysis, their magnitudes get amplified due to the non-Hermiticity in the system, a phenomenon we unfold in this work and name amplified magnetic catalysis in non-Hermitian Dirac systems.

In this work, we exclusively considered the subcritical strengths of the NN Coulomb and on-site Hubbard repulsions (Table~\ref{tab:citicalvalues}). Nonetheless, our conclusions on the NH amplification of magnetic catalysis can be extended to the parameter regime where these two interactions are strong enough to nucleate CDW and AFM orders, respectively, even in the absence of any external magnetic field. In that case, there exists a finite mass gap in the spectrum of Dirac fermions in pristine condition, which increases with the application of external strong perpendicular magnetic fields as $B^2$ in both Euclidean~\cite{magcatagraphene:4, HF:1} and hyperbolic~\cite{Bfieldhyperbolic:3} systems, where $B$ is the magnetic field strength. Introduction of the non-Hermiticity is expected to further amplify the corresponding magnitude of the gap without affecting its characteristic $B^2$ scaling. Here we do not show such expected outcomes explicitly.

Due to the formation of sharp Landau levels at finite energies in Euclidean Dirac systems, subject to external magnetic fields, there exists a universal aspect of the scaling behavior for the interaction induced mass gap. Firstly, notice that the energy of the $n$th Landau level for planar relativistic system scales as $E_n \sim v_{_{\rm F}}\sqrt{n B}$~\cite{grapheneQH:3}, where $v_{_{\rm F}}$ is the Fermi velocity, which can be verified from the linear scaling of $E^2_1$ with $\beta$ shown in Fig.~\ref{fig:Honeyscaling}(a). As non-Hermiticity only scales all the finite eigenvalues by a factor of $\sqrt{1-\alpha^2}$, such a linear scaling between $E^2_1$ and $\beta$ holds even in NH systems, with, however, the slope of the linear fit being $\alpha$-dependent. Next, we note that when the strength of the NN Coulomb repulsion is $V=V_c$, where $V_c$ is the $\alpha$-dependent critical interaction strength for the CDW in the absence of any magnetic field (see Table~\ref{tab:citicalvalues}), $\delta_{\rm CDW} \sim \sqrt{B}$~\cite{magcatagraphene:4, HF:1}, which can be seen from the linear scaling of $\delta^2_{\rm CDW}$ with $\beta$ shown in Fig.~\ref{fig:Honeyscaling}(b) in Hermitian ($\alpha=0$) and NH (finite $\alpha$) systems. Similarly, when $U=U_c$ (also an $\alpha$-dependent quantity), we find $\delta^2_{\rm AFM} \sim \beta$, which is also shown in Fig.~\ref{fig:Honeyscaling}(b) for $\alpha=0$ and a few finite values of $\alpha$ for which the values of $U_c$ are also quoted in Table~\ref{tab:citicalvalues}.

Fascinatingly, all the numerically computed values of $\delta^2_{\rm CDW}$ and $\delta^2_{\rm AFM}$ when compared with $\beta$, fall roughly on a single straight line. This observation confirms that the quantum phase transitions from a Hermitian and NH Dirac semimetal to either one of the two commuting class masses belong to the same universality class in the mean-field limit. As the mean-field analysis ignores quantum fluctuations, we expect that the universality classes associated with the quantum phase transitions for the CDW and AFM orders to belong to distinct universality classes, which, however, should be identical in Hermitian and NH systems, as recently predicted from field theoretical renormalization group calculations~\cite{NHDirac:1}. Finally, we compute the universal ratio $C(\alpha)=E_1/\delta_{x}$ for $x={\rm CDW}$ and ${\rm AFM}$, which is a monotonically decreasing function of $\alpha$, with $C(0) \approx 8.0$, as shown in Fig.~\ref{fig:Honeyscaling}(c). By contrast, the field theoretic prediction yields $C(0) = 5.985$~\cite{magcatagraphene:4}. We believe that such a discrepancy stems from the fact that here the ratio is computed from lattice-based numerical simulations where the magnetic length $\ell_{B} \sim$ a few lattice spacing ($a$), whereas the field theoretic prediction holds for $\ell_B \gg a$. Nonetheless, we find excellent agreement with the predicted scaling $C(\alpha)/C(0) = \sqrt{1-\alpha^2}$, as shown in the inset of Fig.~\ref{fig:Honeyscaling}(c). Such a scaling of the universal ratio with $\alpha$ follows from the fact that $E_1 \equiv E_1(\alpha)=\sqrt{1-\alpha^2} \; E_1(0)$, originating from $v_{_{\rm F}} \equiv v_{_{\rm F}}(\alpha)=\sqrt{1-\alpha^2} \; v_{_{\rm F}}(0)$~\cite{NHcatalysis:1, NHDirac:1}.

For spinful fermions, we here neglect the Zeeman coupling between the electronic spin and the external magnetic fields. Such a coupling confines the AFM order within the easy-plane, perpendicular to the direction of the external magnetic field and at the same time yields easy-axis ferromagnetism, pointing in the external field direction~\cite{magcatagraphene:3, magcatagraphene:6, magcatagraphene:7}. This way the AFM and ferromagnetic orders mutually anticommute with each other and yields the maximal gain in condensation energy. In the future, it will be worthwhile to incorporate the Zeeman coupling in our self-consistent analysis, which, however, is expected not to affect the proposed NH amplification of magnetic catalysis for commuting class masses as the order-parameter for the ferromagnetic order $\hat{{\mathcal O}}_{\rm FM}=\sigma_3 \otimes \; {\rm diag}. ({\bf I}, {\bf I})$ commutes with the total free-fermion NH operator $\sigma_0 \otimes \hat{h}_{\rm NH}$ and anticommutes with $\hat{{\mathcal O}}_{\rm AFM}$ in both Hermitian and NH systems.

Although here we have restricted ourselves to the NH amplification of magnetic catalysis at zero-temperature, certain qualitative statements can be made regarding the consequence of this mechanism at finite temperatures. Recently, it has been shown that the scaling of the associated transition temperature ($t_c$) of a continuous or second-order phase transition with the external magnetic fields closely follows that of the ordering amplitude at zero-temperature for Euclidean Dirac systems~\cite{magcatafinitet:1, magcatafinitet:2}. We believe that such a conclusion should also hold in hyperbolic Dirac liquids, subjected to external magnetic fields at finite temperatures. Therefore, we expect the NH Dirac systems to feature a higher transition temperature than the Hermitian systems and the scaling of $t_c$ with the magnetic field and non-Hermiticity for subcritical strengths of interactions should qualitatively be similar to the ones shown in Figs.~\ref{fig:5_CDWOrderVaryB} and~\ref{fig:8_AFMOrderVaryB}.

It should be noted that the CDW breaks only a discrete symmetry and displays a true long range order at finite temperature, thereby featuring a genuine finite temperature continuous phase transition. The AFM order, on the other hand, is accompanied by two (one) massless Goldstone modes when we ignore (take into account) the Zeeman coupling~\cite{goldstone:1, goldstone:2}. Hence, the AFM does not show any true long-range order at finite temperatures~\cite{MarminWagner:1, MarminWagner:2, MarminWagner:3} and the associated transition temperature, obtained from a mean-field calculation, signifies only a crossover temperature. But, when the AFM order gets confined to the easy-plane due to Zeeman coupling, the requisite U(1) symmetry of the order parameter allows an algebraic order to set in at a lower temperature (in comparison to the mean-field transition temperature) via the Kosterlitz-Thouless transition at least in Euclidean systems~\cite{KosterlitzThouless:1}, which can manifest in the scaling of the longitudinal resistivity with the magnetic field~\cite{KosterlitzThouless:2}. However, the possibility of such a transition on a curved hyperbolic plane remains unknown so far.

Our conclusions regarding the NH amplification of magnetic catalysis should remain unaffected in the presence of inhomogeneous magnetic fields of arbitrary profile as long as the system encloses a finite number of magnetic flux quanta, for the following reasons. The DOS near to the zero-energy is determined by the flux enclosed by the system, which is not affected by non-Hermiticity. The rest of the spectrum, which loses the Landau quantization even in Euclidean Dirac systems, gets squeezed toward zero-energy by a factor of $\sqrt{1-\alpha^2}$. Then our analytical arguments in favor of the NH amplification for commuting class masses from Sec.~\ref{sec:NHamplMagCata} remain equally operative. However, under this circumstance we expect the local CDW and AFM orders, for example, to closely follow the profile of the external inhomogeneous magnetic fields on both Euclidean and hyperbolic Dirac systems, as previously found in the Hermitian limit~\cite{HF:1, Bfieldhyperbolic:3, inhomogeneous:1}.

It is worth noting that Dirac materials can foster another type of magnetic field that unlike the real one, preserves the time-reversal symmetry, and is named \emph{axial} magnetic field. On graphene's honeycomb lattice it points in the opposite directions at two inequivalent valleys in the hexagonal Brillouin zone~\cite{graphene:RMP} and results from specific modulations of NN hopping amplitude. The Aharonov-Casher index theorem remains equally operative for axial magnetic field at least in graphene~\cite{axialcatalysis:1} and the resulting manifold of zero-energy states can catalyze the formation of various orders at half-filling for sufficiently weak interactions~\cite{axialcatalysis:1, axialcatalysis:2, axialcatalysis:3, axialcatalysis:4, axialcatalysis:5, axialcatalysis:6, axialcatalysis:7, HF:2, HF:4}, a phenomenon known as axial magnetic catalysis. Noticeably, time-reversal symmetric axial magnetic fields have been realized experimentally in strained or buckled graphene~\cite{axialExp:1, axialExp:2, axialExp:3}. Even though exact modulations of the NN hopping amplitude in the tight-binding model that can produce such axial magnetic fields on hyperbolic Dirac systems remain unknown so far, the entire discussion in NH systems from Sec.~\ref{sec:nhfree} is nonetheless applicable therein. Hence, based on the current work, we can predict a NH amplification of axial magnetic catalysis of commuting class masses in strained Euclidean and hyperbolic Dirac materials, which we address in a forthcoming publication.

In this work, we restricted ourselves to a subset of all the Dirac masses that fully anti-commute with $\hat{h}_0$ (yielding gapless Dirac fermions in the continuum limit with the signature linearly vanishing DOS in two dimensions), but commutes with $\hat{h}_{\rm mass}$ [Eq.~\eqref{eq:NHGeneral}] such that they anticommute with the full NH operator $\hat{h}_{\rm NH}$, featuring a collection of massless NH Dirac fermions. In the future, it will be worthwhile to scrutinize the fate of the magnetic catalysis mechanism for the mass orders that anticommute with $\hat{h}_{\rm mass}$, hence named \emph{anticommuting class masses}, which, however, commute with the anti-Hermitian component of $\hat{h}_{\rm NH}$. Unfortunately, we cannot address this question within the self-consistent mean-field approximation, employed here. Notice that the effective single-particle operator in the presence of any anticommuting class mass becomes \emph{complex}. Neither those states are sharp nor the filling-factor can be defined consistently in such a system. We then have to rely on exact diagonalization or quantum Monte Carlo simulations~\cite{NHDirac:2} or Green's function method~\cite{NHGreens} to address this issue, which is left as an open problem at this time.

The challenges involved in any experimental demonstration of our concrete theoretical predictions stems from the requisite non-reciprocal hopping patterns, sourcing the non-Hermiticity in the system (Fig.~\ref{fig:NHpeierlssub}). Still, designer electronic materials~\cite{axialExp:3, designer:1, designer:2, designer:3} and cold atomic setups~\cite{optical:1, optical:2}  are two promising platforms where our predicted NH amplification of magnetic catalysis for commuting class masses can, in principle, be observed. While quantum honeycomb lattices have already been created on designer material platforms~\cite{axialExp:3, designer:3}, the $\{10,3 \}$ and $\{ 14,3\}$ hyperbolic tessellations therein can, in principle, be created by growing its substrate on a suitable material with a different thermal expansion coefficient, such that under thermal cooling, the desired negatively-curved substrate is generated, which then can be decorated by the corresponding lattice sites. In cold atomic setups, honeycomb lattice has already been emulated~\cite{opticalgraphene:1, opticalgraphene:2}, and on this platform hyperbolic tessellations can possibly be achieved by suitable arrangements of the laser traps or optical tweezers. Application of external magnetic fields to designer Dirac system should be a relatively straightforward task. On optical lattices the orbital coupling of external magnetic field can be emulated via the coupling of neutral fermions with synthetic gauge fields~\cite{syntheticgauge:1, syntheticgauge:2}.

Notice that the desired non-Hermiticity in the system, yielding a NH operator that features an all-real eigenvalue spectrum over an extended NH parameter regime, results from non-reciprocal hopping that, however, display a staggered pattern between the sites from two sublattices (Fig.~\ref{fig:NHpeierlssub}). This observation suggests that such a non-Hermiticity should result from a balanced gain and loss along the paths connecting the NN sites, which possibly makes the proposed NH model operators realizable in designer materials and optical lattices. Furthermore, given that recent time has witnessed some breakthroughs in achieving NH couplings on optical lattices~\cite{NHoptLatt:1}, we expect that the current work will stimulate new research efforts to experimentally unfold the fascinating phenomenon of NH amplification of magnetic catalysis. On both the material platforms, the strengths of local interactions have already been controlled to a large degree to observe a number of ordered states therein. Therefore, even though it remains challenging, the experimental observation of our predictions should be within the capacity of currently available tools and facilities in laboratories.

\acknowledgments 

This work was supported by NSF CAREER Grant No.\ DMR-2238679 of B.R. We are thankful to Vladimir Juri\v ci\' c for critical reading of the manuscript. Portions of this research were conducted on Lehigh University's Research Computing infrastructure partially supported by NSF Award No.~2019035.

\section*{Data availability}

Numerical codes and data used and generated in this work are available in Ref.~\cite{datacode}.


\end{document}